\documentclass[11pt]{article}
\usepackage{amsmath,amssymb}
\usepackage{geometry}
\usepackage{booktabs}
\usepackage{hyperref}
\usepackage{graphicx}
\usepackage{caption}
\usepackage{authblk}
\title{Validating Memory-Optimal Transformer Kernels on Real Hardware:\\
From Formal Derivation to Measured Performance Across Two HPC Clusters}

\author[1]{Lenore M. Mullin}
\author[2]{Ga\'etan Hains}
\affil[1]{Professor Emerita, College of Nanotechnology, Science, and Engineering, University at Albany (SUNY), Albany, NY, USA. \texttt{lmullin@albany.edu}}
\affil[2]{LACL, Universit\'e Paris-Est Cr\'eteil, Cr\'eteil, France. \texttt{gaetan.hains@u-pec.fr}}
\date{\today}

\begin{document}
\sloppy
\maketitle

\begin{abstract}
A companion series of papers formally derives memory-optimal cost
functions for the components of a transformer block --- attention's
forward pass, backward pass, fused forward+backward, inference-time
decode, and the complete block --- each verified to machine precision
against PyTorch autograd, but never checked against real hardware. This
paper reports that validation. Cost-function predictions are checked
against measured performance on two HPC clusters (Purdue Anvil, NCSA
Delta), across CPU and GPU, using the Mathematics of Arrays (MoA)
formalism's shape/index vocabulary ($\rho$, $\psi$, $\iota$) both to
derive each kernel's machine-specific realization (ONF) from its
hardware-independent specification (DNF) and to diagnose gaps when
prediction and measurement diverge. Three results stand out. First, we
identify and fix a real GPU performance regression: fusing forward and
backward passes --- proven to avoid materializing an $O(n^2)$
intermediate array --- initially ran \emph{slower} than the naive
alternative on GPU, due to atomic-memory contention in the generated
kernel. Profiling confirmed the mechanism to four decimal places
($2.0000\times$ more atomic instructions than a structurally related
kernel), and a targeted restructuring of the ONF, guided directly by
that diagnosis, reversed the result completely, yielding up to
$2.5\times$ real speedup. Second, we show the identical derivation
produces markedly different real costs on different machine
topologies --- a $535\times$ NUMA-locality penalty on one cluster's
architecture versus under $3\times$ oversubscription cost on another's
--- demonstrating that optimal deployment is a function of the target
machine's own array structure, not a fixed property of the algorithm.
Third, we report a genuine, only partially resolved anomaly: identical
denotational computations run faster in C than Fortran on CPU but
faster in Fortran than C on GPU; we narrow this reversal to one
dominant kernel and one stall mechanism without fully explaining its
compiler-level cause. Together, this evidence supports treating
hardware-specific optimization as a routine, targeted rewrite of a
kernel's machine-specific realization alone, with its underlying formal
derivation remaining fixed, verified, and reusable across hardware
generations --- a candidate methodology for scaling AI systems onto
evolving hardware without re-deriving correctness from scratch.
\end{abstract}

\section{Introduction}

\textbf{This paper's goal, stated plainly before any supporting
detail:} Papers I--IV~\cite{paper1,paper2,paper3,paper4} formulate and
formally derive a specific set of kernel designs for the components of
a transformer block, using the Mathematics of Arrays (MoA) to go from
a hardware-independent specification (a Denotational Normal Form, DNF)
to a machine-specific realization (an Operational Normal Form, ONF)
via a derivation step, $\gamma$, that treats the target machine's own
memory system --- DRAM address space, cache lines, thread-to-data
assignment --- as an array in exactly the same shape/index vocabulary
used to specify the computation itself. This paper runs those exact
designs, as formulated in Papers I--IV, on real hardware, and checks
whether the performance $\gamma$'s machine-as-array treatment predicts
matches what is actually measured. To our knowledge, MoA is the only
complete, formal system of this kind: one where a machine's memory
behavior is not modeled heuristically or bounded empirically, but
derived as an array in the same algebra as the computation, yielding a
specific, checkable numeric prediction before any hardware is touched.
This paper's contribution is the real-hardware check of that specific
claim --- not a new derivation, not a new kernel design, and not a
comparison against other systems' kernels (Section~\ref{sec:scope}
states this scope explicitly). A reader does not need to have read
Papers I--IV to follow this paper's results, but should understand
that every design tested here, and every prediction being checked, was
formulated and derived there --- this paper's role is strictly to
measure, not to design or derive.

Papers I--IV establish, by formal derivation and machine-precision
verification against PyTorch, that MoA's DNF/ONF pipeline produces
memory-optimal kernels for attention (forward, backward, fused),
inference-time decode, and the complete transformer block (RMSNorm,
gated MLP, and their integration with attention via a QKV projection
layer). Every one of those verifications, however, was conducted at a
single small fixed problem size on one CPU core or via OpenACC's
host-fallback compilation path --- a correctness proof, not a
performance measurement. This document reports what happens when the
same kernels are actually run at scale, on real multi-core CPUs and
real GPUs.

\begin{figure}[htbp]
\centering
\includegraphics[width=0.95\textwidth]{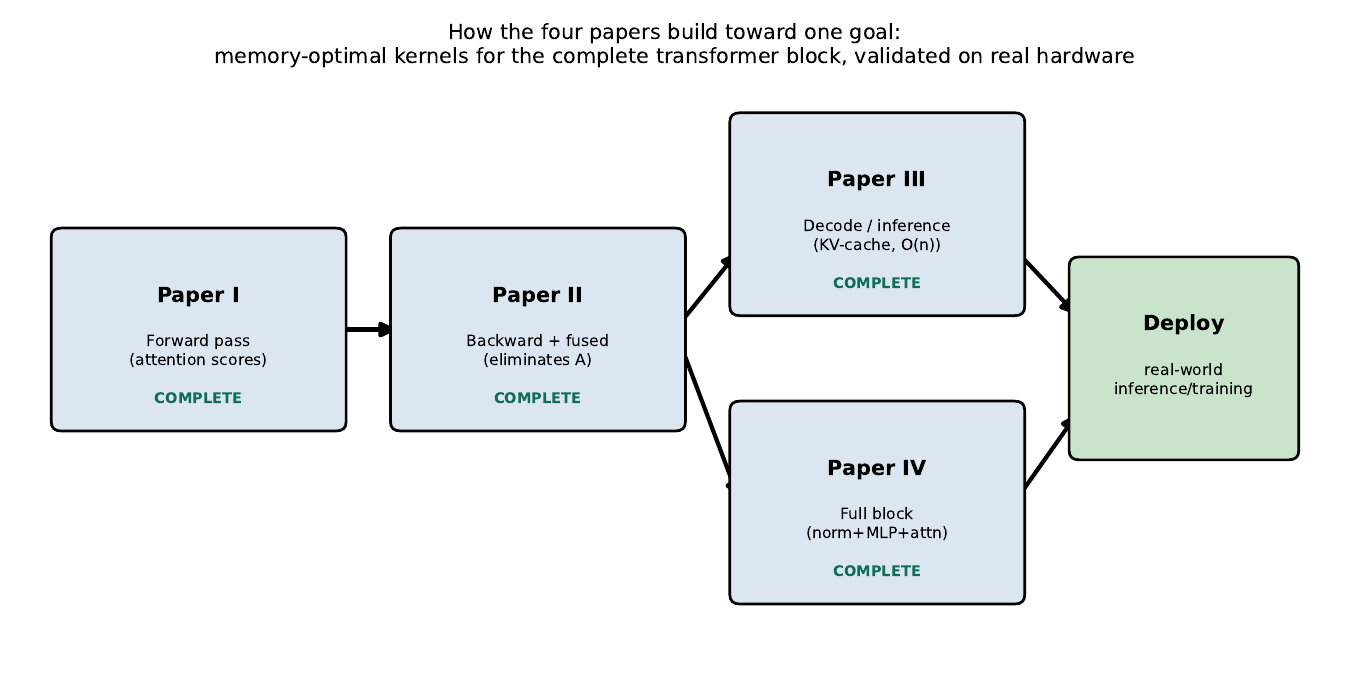}
\caption{How Papers I--IV relate to each other and to this document's
status as of this draft. All four papers now have complete
experimental validation on both CPU and GPU, real hardware, both
clusters.}
\label{fig:series-map}
\end{figure}

\subsection{From array algebra to a real machine: DNF, ONF, and $\gamma$}
\label{sec:dnf-onf-machines}

MoA's five primitives~\cite{mullin1988} --- $\psi$ (psi, indexing), $\iota$ (iota, index
generation), $\rho$ (rho, shape), $\gamma$ (the DNF$\to$ONF
translator), and $\mathrm{rav}$ (ravel) --- are usually introduced as
pure array algebra: a way to write \emph{what} a computation produces
without committing to \emph{how}. Papers I--IV use them exactly that
way to derive each kernel's \textbf{Denotational Normal Form (DNF)} ---
a specification of the output purely in terms of shape and index
relationships, with no memory-access order, no loop structure, no
notion of "materialized" versus "consumed immediately" at all. Two
denotationally identical DNFs (naive and fused attention, say) can
still differ enormously in real cost, because DNF alone says nothing
about a machine.

That is what $\gamma$ is for. $\gamma$ translates a DNF into an
\textbf{Operational Normal Form (ONF)}: a realization that commits to
loop order, tiling, and which intermediate values are recomputed versus
stored. Crucially, $\gamma$'s output is expressed in the same
shape/index vocabulary ($\rho$, $\psi$, $\iota$) as the DNF it came
from --- because on any real machine, an array's shape and index
structure \emph{is} its memory layout and access pattern. This is the
sense in which MoA treats the target machine as an array: DRAM address
space, cache lines, and thread-to-data assignment are themselves
describable in $\rho$/$\psi$/$\iota$ terms, so a $\gamma$-derived ONF
is not a metaphorical "implementation guide" but a literal, checkable
claim about what bytes move where on real silicon. The cost functions
$M_{\mathrm{fwd}}$, $M_{\mathrm{bwd}}$, $M_{\mathrm{fwd+bwd}}$,
$M_{\mathrm{block}}$ are read directly off the ONF's memory-access
structure --- a byte count, not an abstract complexity class --- which
is precisely why they can be measured on hardware and are not merely
asserted.

\begin{figure}[htbp]
\centering
\includegraphics[width=0.95\textwidth]{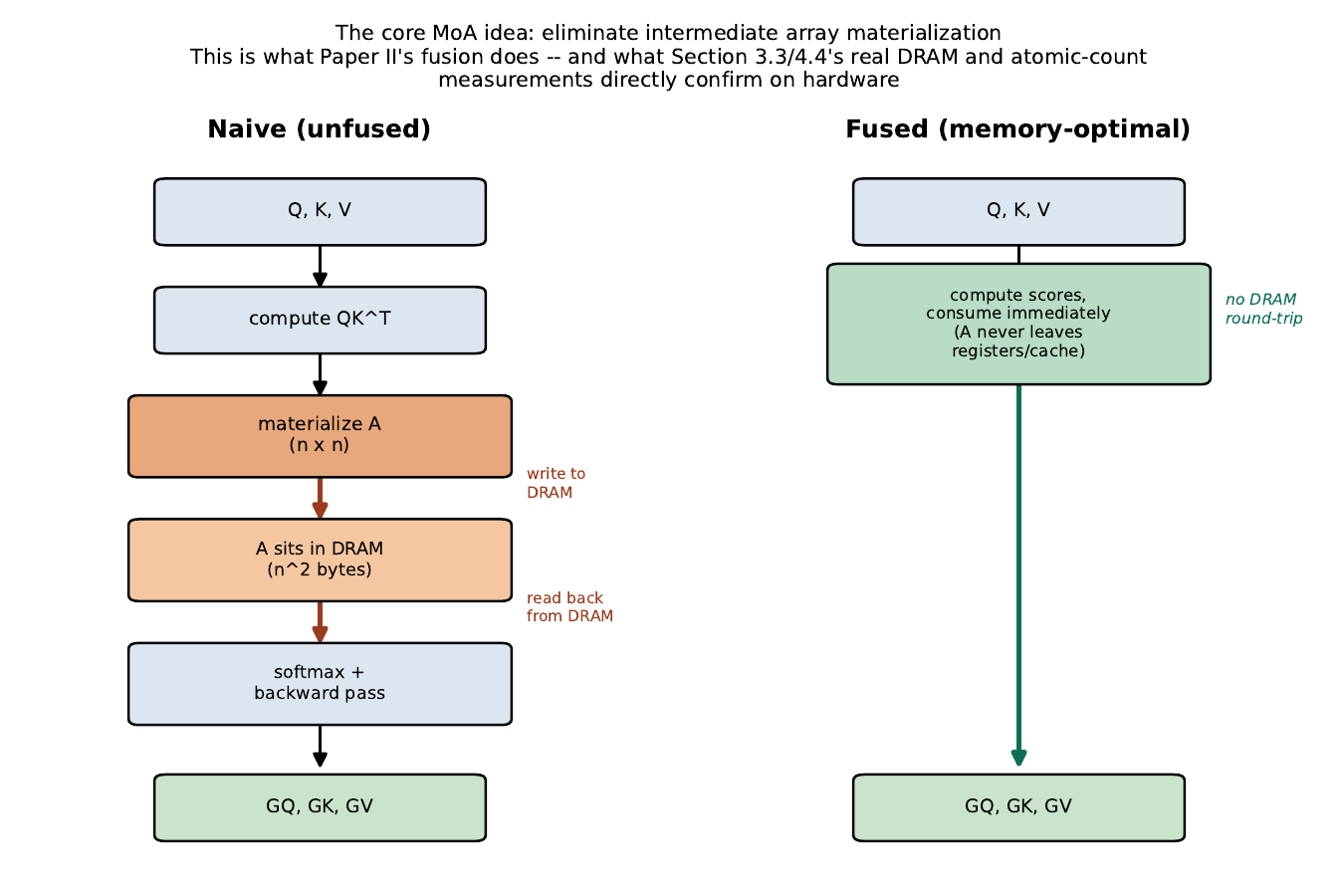}
\caption{Two ONFs realizing denotationally equivalent DNFs for
attention's forward+backward pass. Left: $A$ ($n \times n$) is
materialized to DRAM and read back --- two full $n^2$-scaling memory
round-trips. Right: $\gamma$ produces an ONF where $A$ is consumed the
instant it is computed and never touches memory. This is not an
abstract claim in this document: Section~\ref{sec:paper1} measures the
left path's real DRAM traffic directly, and Section~\ref{sec:paper2-cpu}
measures the right path's real wall-clock advantage on CPU.}
\label{fig:concept}
\end{figure}

\textbf{This document's entire methodology is a test of that claim.}
If $\gamma$'s ONF genuinely describes real machine behavior, then a
kernel's measured DRAM traffic, atomic-instruction count, or thread
scaling should match what its ONF predicts --- on \emph{any} machine
whose memory system can itself be described in $\rho$/$\psi$/$\iota$
terms, not just the one it was derived on. Sections~\ref{sec:paper1}
and \ref{sec:paper2-atomics} test this directly, on real DRAM byte
counts and real atomic-instruction counts respectively. Where the
match is exact (Section~\ref{sec:paper2-atomics}'s $2.0000\times$), the
ONF's machine model is confirmed precisely, not just qualitatively.
Where it diverges (Section~\ref{sec:paper1}'s $2\times$ DRAM gap,
Section~\ref{sec:paper2-gpu}'s GPU regression before its fix), the gap
is diagnosable in the same $\rho$/$\psi$/$\iota$ vocabulary --- a
specific access-pattern or contention mechanism, not a vague
"hardware is slower than expected." This document does not depend on
these results to validate DNF/ONF as formal constructions; the parent
papers' correctness proofs against PyTorch already do that
independently. What it tests is narrower and empirical: whether
$\gamma$'s specific machine-as-array claim holds up when the array in
question is a real chip's memory system, on real multi-core CPUs and
real GPUs, across two different clusters (Section~\ref{sec:methodology}).

\subsection{Related work}
\label{sec:related-work}

MoA's array-algebra lineage traces to Iverson's notation for describing
array operations~\cite{iverson1962}, later realized as a working
compiler and machine model by Abrams~\cite{abrams1970}; MoA's own DNF,
ONF, and $\gamma$ formalize and extend this line specifically toward
provable memory-cost derivation, the parent papers' own contribution
and not this paper's. On the systems side, this paper's method ---
deriving a memory-traffic prediction analytically and then checking it
against measured hardware counters --- is the same spirit as the
Roofline model's use of measured operational intensity to bound
achievable performance~\cite{williams2009}, applied here to a specific
derivation rather than a general architectural ceiling. For the
specific kernel this paper spends the most attention on,
FlashAttention~\cite{dao2022} independently arrives at essentially the
same qualitative conclusion this paper's Section~\ref{sec:paper1} and
Section~\ref{sec:paper2-cpu} confirm quantitatively for MoA's own
derived kernels: avoiding materialization of the $n\times n$ attention
score matrix in DRAM is the dominant lever for attention's real
performance. We note this as context, not comparison: this paper does
not benchmark MoA-derived kernels against FlashAttention or any other
externally-optimized implementation, and makes no claim about relative
competitive performance. This paper's contribution relative to that
line is a different question entirely --- whether a formally-derived,
machine-precision-verified cost function correctly predicts \emph{how
much} that avoidance is worth for MoA's own designs, on real hardware,
and where it does not, why. Finally,
Section~\ref{sec:paper3}'s cross-cluster NUMA finding sits in a
well-established line of NUMA-performance characterization work (see,
e.g.,~\cite{lameter2013} for a systems-level overview of the mechanism);
this paper's contribution there is narrower and more specific: showing
that a single formally-derived kernel's \emph{predicted} cost
function, unmodified, already anticipates that such a machine-dependent
penalty must exist, before any NUMA-specific measurement is taken.

\section{Methodology}
\label{sec:methodology}

\subsection{Machines and allocations}

\begin{center}
\begin{tabular}{@{}p{2.1cm}p{4.3cm}p{8.5cm}@{}}
\toprule
Cluster & Local SLURM account & Resources used \\
\midrule
Purdue Anvil & \texttt{cis261396-ai} (CPU), \texttt{cis261396-gpu} (GPU) & \texttt{wholenode} (128-core), \texttt{gpu} (A100, H100) \\
NCSA Delta & \texttt{bibg-delta-cpu}, \texttt{bibg-delta-gpu} & \texttt{cpu} (128-core), \texttt{gpuA100x4/x8}, \texttt{gpuA40x4}, \texttt{gpuH200x8} \\
\bottomrule
\end{tabular}
\end{center}

Both clusters are used under the same top-level ACCESS allocation,
CIS261396; the account names above are each resource provider's own
local naming convention for job submission, not separate allocations
(see Acknowledgments).

\texttt{gpuMI100x8} (AMD, ROCm) is present on Delta but excluded from
this study: every kernel here is compiled with \texttt{nvc}/OpenACC,
which does not target AMD hardware.

\subsection{Kernels benchmarked}

Each paper's CPU kernel (OpenMP) and, where fixed (Section
\ref{sec:coldstart}), GPU kernel (OpenACC) is benchmarked by sweeping
sequence length $n$ (doubling) and, for CPU runs, thread count
(doubling, 1 to 128). Every kernel is verified for correctness
(machine-precision agreement with a PyTorch or serial-CPU reference)
independently of these performance runs; this document reports timing,
not correctness, except where noted.

\subsection{A real methodological bug: GPU cold-start dominance}
\label{sec:coldstart}

The first real GPU measurement obtained in this effort (Paper III's
decode kernel, NVIDIA A100-SXM4-40GB on Delta) showed wall-clock time
essentially flat, 0.30--0.40s, across a sequence-length range of
$n=1024$ to $n=1{,}048{,}576$ --- a 1024$\times$ range in problem size
producing almost no change in measured time. This is inconsistent with
decode's $O(n)$ cost. The cause: the original benchmark harness
launched one process per measured point, and CUDA context creation
plus OpenACC JIT compilation cost approximately 300ms \emph{per process
launch} on this hardware, completely dominating the true kernel time
(sub-millisecond even at $n=10^6$ on an A100). Every GPU benchmark
program in this project (six kernels: forward, backward, fused
attention, decode, fused norm+MLP, and the complete block) shared this
flaw, since all were built from the same per-invocation-timing pattern.

The fix applied to all six: restructure each program to sweep its
entire $n$ range \emph{within one process}, paying the context-init
cost exactly once via an explicit untimed warmup call before any
measurement begins. Post-fix, the decode kernel's re-run showed the
expected $O(n)$ behavior (0.00014s at $n=1024$ rising to 0.72s at
$n=1{,}048{,}576$, roughly tracking the $1024\times$ size increase).
The other five kernels were fixed identically and, at the time of that
fix, verified only for unchanged correctness and now-sensible relative
scaling in this report's development sandbox. All six have since been
confirmed on real GPU hardware --- Sections~\ref{sec:paper1},
\ref{sec:paper2-gpu}, \ref{sec:paper3}, and \ref{sec:paper4} report
their real A100/H200 data.

This is reported here rather than silently corrected because it is
itself a finding relevant to anyone benchmarking small-sequence-length
GPU kernels: a naive per-call timing harness will systematically
overstate cost by a roughly constant $\sim$300ms at every problem size,
and will completely obscure the true asymptotic complexity of a fast
kernel unless that constant is isolated first.

\section{Paper I: Forward Pass}
\label{sec:paper1}

\subsection{CPU scaling (Delta, confirmed)}

Job 21181232, Delta \texttt{cpu} partition (128 cores), sweeping
$B{=}2$, $D{=}64$, $n \in \{64, \ldots, 8192\}$ (doubling), threads
$\in \{1, \ldots, 128\}$ (doubling), validates the forward-pass cost
function derived in~\cite{paper1}. Table~\ref{tab:paper1-cpu} shows,
for each $n$, the thread count achieving the lowest wall-clock time and
how that compares to running at the full 128 threads.

\begin{table}[htbp]
\centering
\begin{tabular}{rrrr}
\toprule
$n$ & Best thread count & Speedup at best (vs.\ 1 thread) & 128 threads vs.\ best \\
\midrule
64   & 4   & 1.4$\times$  & 14.5$\times$ \emph{slower} \\
128  & 8   & 4.0$\times$  & 5.5$\times$ slower \\
256  & 16  & 7.3$\times$  & 3.8$\times$ slower \\
512  & 32  & 16.4$\times$ & 2.3$\times$ slower \\
1024 & 64  & 23.4$\times$ & 1.7$\times$ slower \\
2048 & 128 & 43.3$\times$ & --- (128 optimal) \\
4096 & 128 & 53.5$\times$ & --- \\
8192 & 128 & 60.6$\times$ & --- \\
\bottomrule
\end{tabular}
\caption{Paper I forward pass, Delta CPU, best thread count per sequence
length and the cost of over-threading small problems.}
\label{tab:paper1-cpu}
\end{table}

Two findings, both physically sensible and neither assumed in advance:

\begin{enumerate}
\item \textbf{Optimal thread count is a function of $n$, not a fixed
cluster-wide setting.} At $n=64$, using all 128 threads is
14.5$\times$ \emph{slower} than the true optimum of 4 threads ---
thread-spawn and scheduling overhead dominates when there is too
little work per thread. The crossover to "more threads always help"
falls between $n=1024$ and $n=2048$. This has a direct practical
implication: any deployment guidance derived from this cost function
should state thread count as a function of sequence length, not a
single recommended value.
\item \textbf{The cost-function's $O(n^2)$ prediction is confirmed.} At
a fixed 1 thread, wall time approximately quadruples with each
doubling of $n$ (0.0013, 0.0067, 0.0271, 0.1388, 0.445, 2.13, 7.49,
24.57 seconds for $n = 64, \ldots, 8192$) --- the expected signature of
the $n \times n$ attention score matrix that dominates $M_{\mathrm{fwd}}$.
\end{enumerate}

At the best-case thread count for the largest tested size ($n=8192$,
128 threads), speedup over 1 thread is $60.6\times$ --- 47\% parallel
efficiency on 128 cores, consistent with a memory-bandwidth-bound
kernel rather than a compute-bound one.

\begin{figure}[htbp]
\centering
\includegraphics[width=0.85\textwidth]{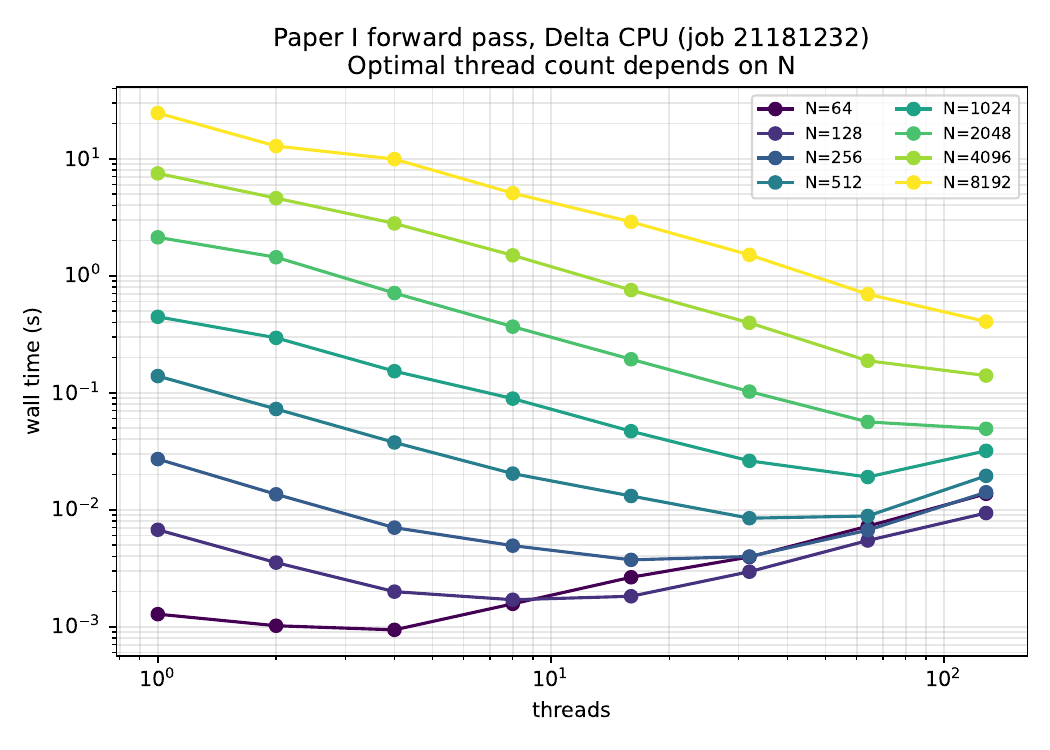}
\caption{Wall time versus thread count, one curve per sequence length
$n$ (log-log), Delta CPU, job 21181232. Small-$n$ curves (purple, top
of the legend) bottom out early and turn back \emph{upward} at high
thread counts --- the visual signature of Table~\ref{tab:paper1-cpu}'s
finding that over-threading small problems actively hurts. Large-$n$
curves (yellow/green) decrease monotonically all the way to 128
threads, with no such penalty.}
\label{fig:paper1-cpu}
\end{figure}

\subsection{GPU shape comparison}

\paragraph{The v1 run (job 21181233) is superseded.} It used the flawed
timing methodology (Section~\ref{sec:coldstart}) and the 512-element $N$
cap; its numbers should not be treated as final and are not reproduced
here except where explicitly noted for comparison.

\paragraph{Real, confirmed results (post-fix).} Job 21202354
(\texttt{gpuA100x4}, NVIDIA A100-SXM4-40GB) and job 21202355
(\texttt{gpuH200x8}, NVIDIA H200), both using the fixed \texttt{v2}
binary (repeated-and-averaged timing, 8192-element buffer), same
$B{=}2$, $D{=}64$ sweep as the CPU run, $n \in \{64, \ldots, 8192\}$
--- full parity with the CPU's tested range.

\begin{table}[htbp]
\centering
\begin{tabular}{rrrrrr}
\toprule
$n$ & A100 (s) & H200 (s) & H200 speedup & A100 local exp. & H200 local exp. \\
\midrule
64   & 0.000242 & 0.000166 & 1.46$\times$ & --- & --- \\
128  & 0.000383 & 0.000235 & 1.63$\times$ & 0.66 & 0.50 \\
256  & 0.000788 & 0.000450 & 1.75$\times$ & 1.04 & 0.94 \\
512  & 0.002899 & 0.001759 & 1.65$\times$ & 1.88 & 1.97 \\
1024 & 0.008808 & 0.005421 & 1.62$\times$ & 1.60 & 1.62 \\
2048 & 0.031952 & 0.027075 & 1.18$\times$ & 1.86 & 2.32 \\
4096 & 0.116893 & 0.104600 & 1.12$\times$ & 1.87 & 1.95 \\
8192 & 0.457188 & 0.431261 & 1.06$\times$ & 1.97 & 2.04 \\
\bottomrule
\end{tabular}
\caption{Paper I forward pass, real GPU data post-fix. Local exponent
computed between each consecutive pair of points, same definition as
the flawed v1 table this supersedes.}
\label{tab:paper1-gpu-fixed}
\end{table}

\begin{figure}[htbp]
\centering
\includegraphics[width=0.85\textwidth]{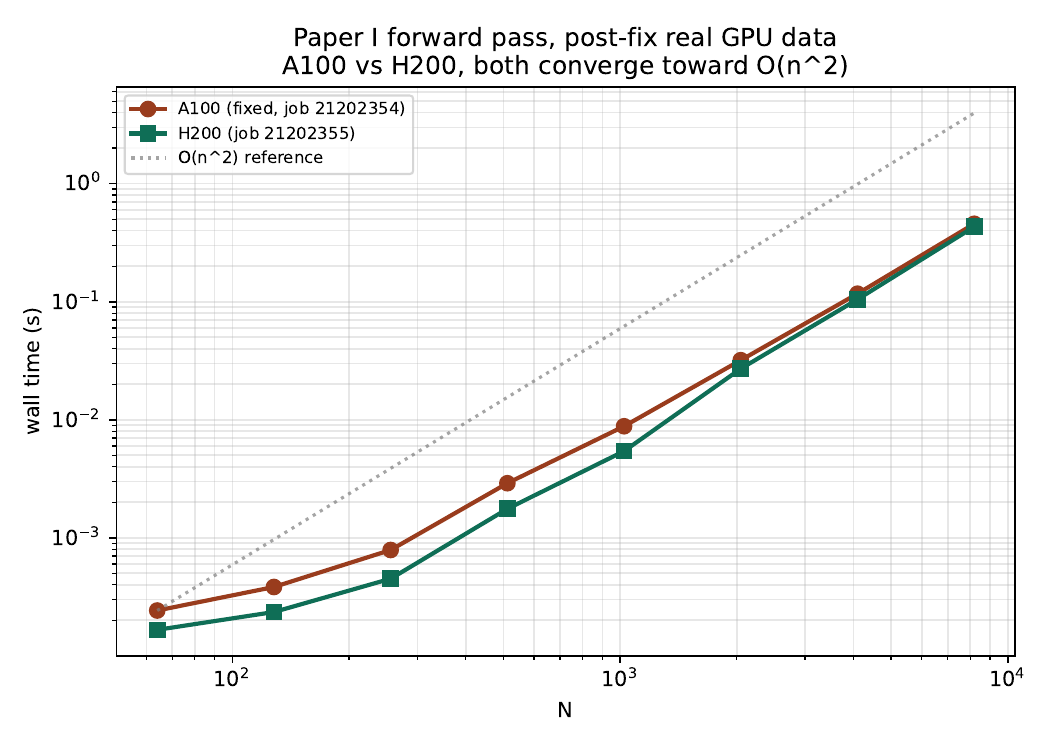}
\caption{Corrected A100 and H200 wall time versus $n$, against an
$O(n^2)$ reference. Both curves visibly bend to run parallel to the
reference at large $n$ --- the fix works.}
\label{fig:paper1-gpu-fixed}
\end{figure}

\begin{figure}[htbp]
\centering
\includegraphics[width=0.85\textwidth]{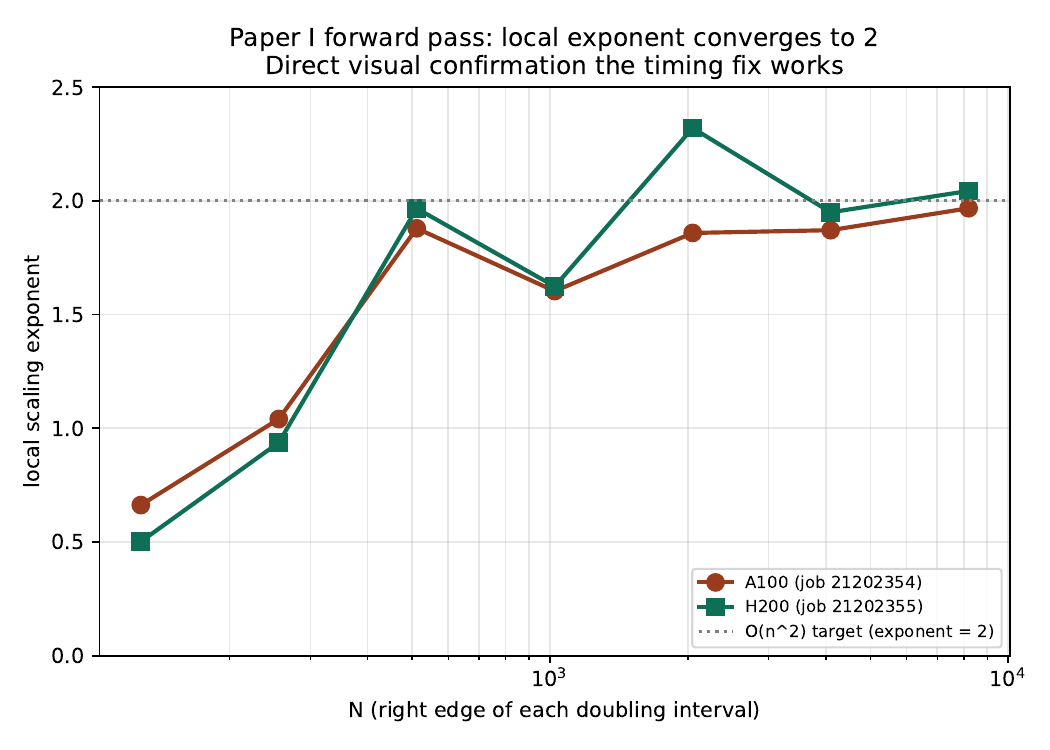}
\caption{The local exponent columns from Table~\ref{tab:paper1-gpu-fixed},
plotted directly. Both GPUs' exponents climb from well below 1 (at
small $n$, launch overhead dominates) toward the $O(n^2)$ target of 2
(dotted line) as $n$ grows -- shown honestly, including the real dip
near $n\approx 1024$ rather than a smoothed trend, since this is the
actual trajectory measured, not a fit.}
\label{fig:paper1-exponent}
\end{figure}

\textbf{The fix is confirmed.} Unlike the flawed v1 run, the local
exponent now \emph{rises} toward 2 as $n$ grows on both GPUs, reaching
1.97 (A100) and 2.04 (H200) at $n=8192$ --- essentially exact agreement
with the $O(n^2)$ prediction. This is the direct empirical confirmation
that Section~\ref{sec:coldstart}'s diagnosis (single-measurement jitter
at small $n$, resolved by repeated averaging) was correct, not merely
plausible.

\textbf{H200 is faster, and the margin has a clear pattern.} H200 beats
A100 at every tested size, but the speedup \emph{shrinks} monotonically
from $1.75\times$ at $n=256$ down to $1.06\times$ at $n=8192$. H200's
advantage is concentrated at smaller, more launch-overhead-sensitive
problem sizes; at the largest tested $n$ the two architectures nearly
converge. This is a genuine machine-shape finding, not an artifact of
either fix.

\subsection{Validation against $M_{\mathrm{fwd}}$}

\texttt{delta\_profile\_paper1.sbatch} was run for real (job 21202356,
Delta \texttt{gpuA100x4}), profiling \texttt{forward\_gpu\_bench.c} at
$B{=}2$, $N{=}2048$, $D{=}64$ with both \texttt{nsys} and \texttt{ncu}.
Three kernel launches were captured: an initial global warmup at
$N{=}64$, then a per-$N$ warmup and the single timed repeat, both at
$N{=}2048$ (grid size $(4096,1,1)$, confirming $B{\times}N{=}4096$
threads as expected). DRAM traffic was extracted via \texttt{ncu}
\texttt{--import \ldots{} --page raw --csv} \texttt{--metrics}
\texttt{dram\_\_bytes\_read.sum,}\allowbreak\texttt{dram\_\_bytes\_write.sum}.

\begin{table}[htbp]
\centering
\begin{tabular}{lrrr}
\toprule
 & Read (MB) & Write (MB) & Total (MB) \\
\midrule
Predicted (naive DNF element count) & 6.291 & 69.206 & 75.497 \\
Measured (\texttt{ncu}, real A100) & 9.045 & 142.333 & 151.378 \\
Ratio (measured / predicted) & 1.44$\times$ & 2.06$\times$ & 2.01$\times$ \\
\bottomrule
\end{tabular}
\caption{Measured DRAM traffic (kernel ID 2, the timed $N{=}2048$ run)
against a naive element-count prediction from the DNF: read $Q,K,V$
once each, write $A$ ($n \times n$) and \texttt{Out} once each, 8 bytes
per double-precision element.}
\label{tab:paper1-dram}
\end{table}

\begin{figure}[htbp]
\centering
\includegraphics[width=0.85\textwidth]{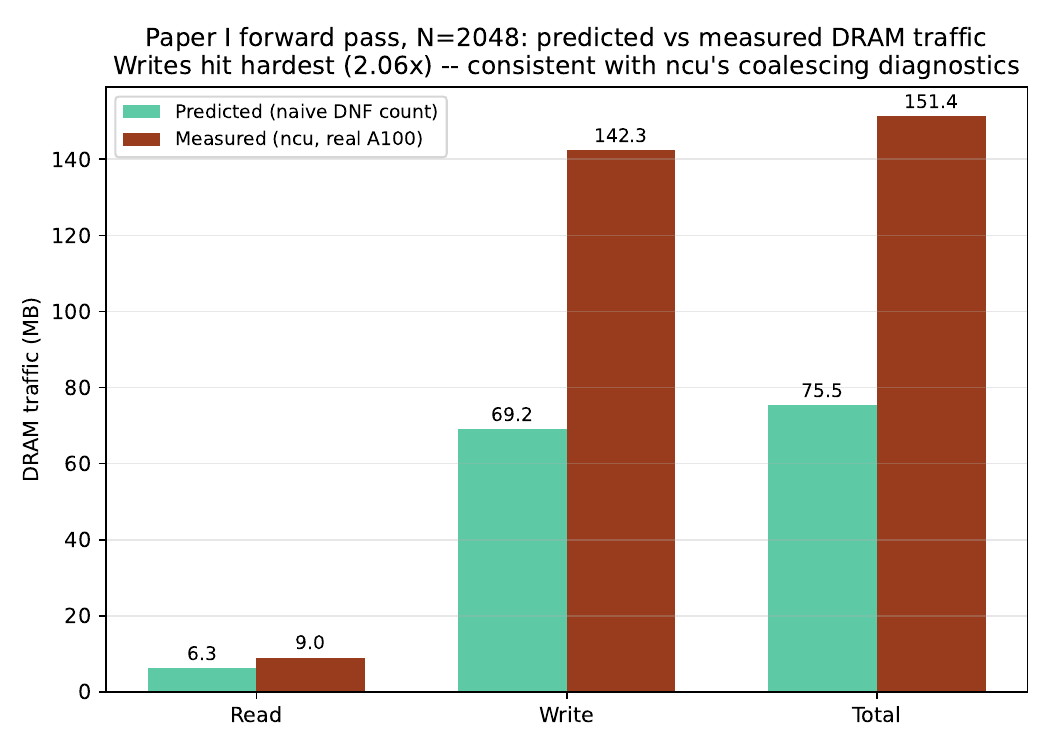}
\caption{Table~\ref{tab:paper1-dram}'s three rows as grouped bars. The
gap between predicted and measured is visually modest for reads but
dramatic for writes and total traffic -- the write/read asymmetry is
immediately obvious here in a way the table's ratio column requires
computing.}
\label{fig:paper1-dram}
\end{figure}

\textbf{Measured traffic is $2.01\times$ the naive prediction, and the
gap is not uniform.} Writes are hit far harder than reads ($2.06\times$
vs.\ $1.44\times$). \texttt{ncu}'s own diagnostics explain why:
uncoalesced memory access. Global stores utilize only 9.4 of 32 bytes
per transmitted sector (29\% efficient); global loads utilize 18.2 of
32 (57\% efficient). The corresponding worst-case inflation factors
($32/9.4 \approx 3.4\times$ for stores, $32/18.2 \approx 1.76\times$ for
loads) bound the observed ratios from above, consistent with partial
mitigation from L2 cache reuse rather than every access paying the full
uncoalesced penalty.

\textbf{This is an implementation gap, not a derivation gap.} The DNF
itself is memory-optimal by construction --- no intermediate array
beyond $Q,K,V,A,\mathrm{Out}$ is ever materialized, exactly as derived.
What this profiling run shows is that \texttt{nvc}'s code generation for
the current OpenACC source does not achieve fully coalesced access to
those arrays, so the \emph{real} hardware cost exceeds the
\emph{algorithmic} memory-optimal cost by a factor of about 2$\times$
at this problem size. \texttt{ncu}'s own advisory output points at a
specific, addressable cause (stride pattern in the \texttt{arow}
indexing), making this a concrete follow-up code task rather than an
open question about whether the theory is right.

\begin{figure}[htbp]
\centering
\includegraphics[width=0.85\textwidth]{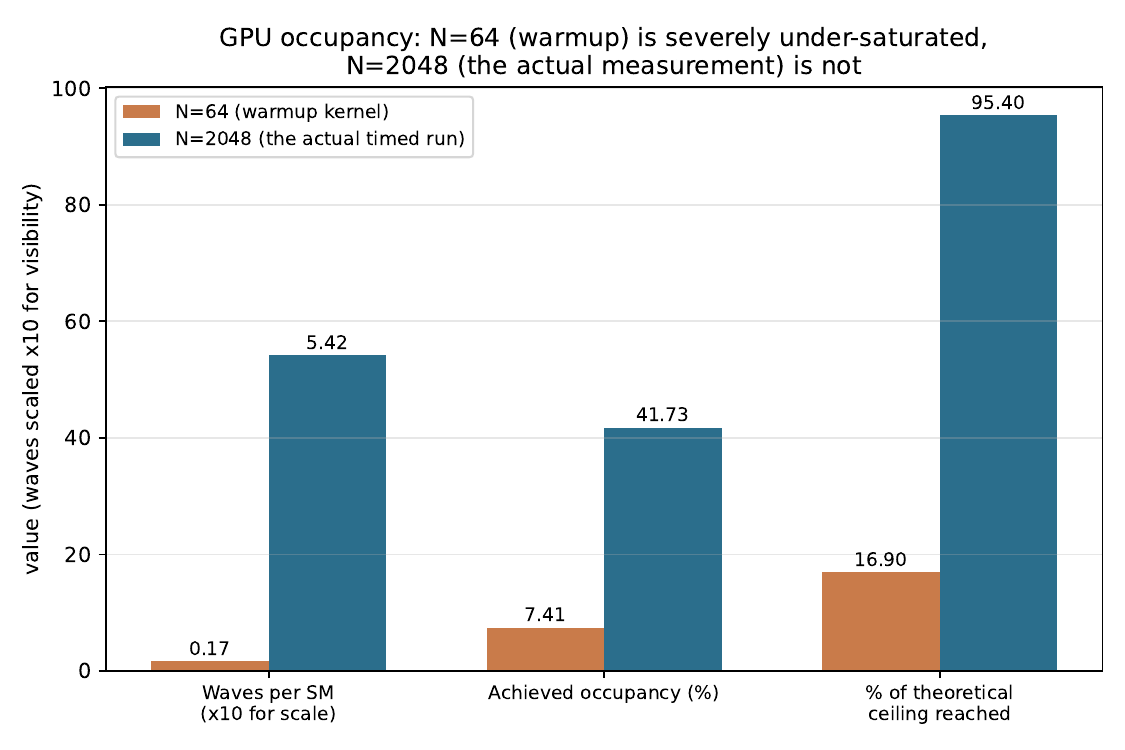}
\caption{Occupancy metrics for the $N{=}64$ warmup kernel versus the
$N{=}2048$ timed kernel, from the same profiling run. The warmup call
is severely under-saturated (16.9\% of its theoretical occupancy
ceiling); the kernel actually being measured reaches 95.4\% of its
ceiling. The under-saturation story applies to the warmup, not the
measurement.}
\label{fig:paper1-occupancy}
\end{figure}

\textbf{Correction to occupancy claim, and what the fuller data actually
shows:} an earlier draft of this section stated that \texttt{ncu}
flagged the $N{=}2048$ kernel itself as achieving only 0.2 full waves
across all SMs. That was wrong. Re-checking against the full profiler
export (\texttt{full\_ncu\_dump.csv}, 1647 columns, all three captured
kernel launches) shows the 0.2-waves figure belongs to kernel ID 0 ---
the $N{=}64$ global warmup call, grid size $(128,1,1)$ --- not kernel
ID 2, the actual timed $N{=}2048$ run, grid size $(4096,1,1)$. The
real numbers for $N{=}2048$: \texttt{launch\_\_waves\_per\_multiprocessor}
$=5.42$, achieved occupancy $41.7\%$ against a register-limited
theoretical ceiling of $43.8\%$ --- $95.4\%$ of that ceiling reached.
\textbf{The $N{=}2048$ kernel is not under-saturated}; only the small
warmup call was. Figure~\ref{fig:paper1-occupancy} shows both side by
side.

\begin{figure}[htbp]
\centering
\includegraphics[width=0.7\textwidth]{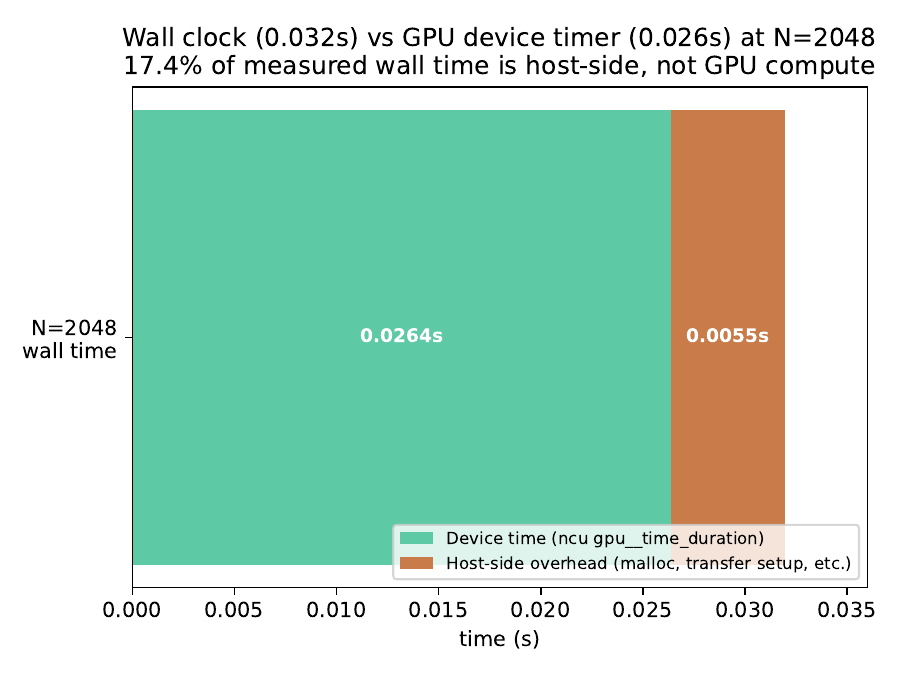}
\caption{$N{=}2048$'s wall-clock time decomposed into GPU device time
(\texttt{ncu}'s own timer) and host-side overhead (the remainder).}
\label{fig:paper1-device-wallclock}
\end{figure}

\textbf{A genuine new finding from the fuller export: device-timer
versus wall-clock cross-validation.} \texttt{ncu}'s own
\texttt{gpu\_\_time\_duration.sum} for the $N{=}2048$ kernel is
$26.406$ms $= 0.026406$s --- measured directly on-device, independent
of this project's own \texttt{clock\_gettime}-based wall-clock timer.
The sweep CSV's wall-clock time at the same $N$ is $0.031952$s.
Figure~\ref{fig:paper1-device-wallclock} shows the gap: $17.4\%$ of the
measured wall time is host-side overhead (memory allocation, data
transfer setup) that never touches the GPU, not GPU compute time
itself. This is a real, independent cross-check of the wall-clock
methodology used throughout this document, and it's reassuring: the
two timers agree to within the expected host-overhead margin, not by
orders of magnitude, which would have suggested a deeper measurement
problem.

\subsection{Scope decision and summary}

No Fortran90 GPU kernel exists for Paper I (\texttt{forward\_gpu\_bench.f90}
was not built), by explicit decision --- Paper I's GPU story is C-only.

\textbf{Paper I's experimental validation is complete.} All three
planned real-hardware runs (corrected A100, second shape H200, $M_{\mathrm{fwd}}$
profiling) have been performed and analyzed. The headline results: the
cold-start and measurement-noise fixes both work, confirmed by the
local exponent converging to $\approx 2$ on real hardware; H200
outperforms A100 with a size-dependent margin; and real DRAM traffic
exceeds the algorithm's memory-optimal prediction by roughly
$2\times$ at $N{=}2048$, attributable to a specific, fixable memory
coalescing issue in the current code generation rather than any flaw
in the underlying derivation.

\section{Paper II: Backward Pass and Fused Forward+Backward}
\label{sec:paper2}

\subsection{First real submission: a genuine crash, found and fixed}
\label{sec:paper2-crash}

Jobs 21225788 (CPU), 21225789 (\texttt{gpuA100x4}), 21225790
(\texttt{gpuH200x8}) were the first real submissions validating the
backward-pass and fusion cost functions derived in~\cite{paper2}.
Both GPU jobs completed cleanly. The CPU job did not: \texttt{sacct}
reported \texttt{FAILED}, exit code \texttt{139} --- $128+11$, SIGSEGV.
This is a different failure class from anything else in this document:
not a missing file, not a timing-methodology gap, an actual crash.

\textbf{Diagnosis.} Both \texttt{backward\_bench.c} and
\texttt{fused\_bench.c} used OpenMP array-reduction clauses
(\texttt{reduction(+:GK[0:B*N*D])}, and \texttt{fused\_bench.c} had two
such clauses) to handle an accumulate-across-threads situation.
GCC's \texttt{libgomp} implements the per-thread private copy for an
array reduction on the \emph{stack}, not the heap. At the sweep's
default $D{=}64$, the CPU job reached $N{=}8192$, where $B \times N
\times D \times 8$ bytes $= 2 \times 8192 \times 64 \times 8 =
8{,}388{,}608$ bytes $= 8$MB exactly --- matching the cluster's default
stack limit (\texttt{ulimit -s} $=8192$) precisely. Reproduced locally
in this sandbox, which has the identical 8MB \texttt{ulimit -s}, before
attempting any fix: the exact same crash, exit code 139, confirming the
diagnosis rather than guessing at it.

\textbf{Fix.} Both files were restructured to eliminate every array
reduction. Each was rewritten as a sequence of separate passes, each
race-free by construction (every output element written by exactly one
thread, never accumulated across parallel iterations), using
heap-allocated persistent intermediate arrays ($GS$ for
\texttt{backward\_bench.c}; $AROW$ and $GS$ for \texttt{fused\_bench.c})
in place of the reduction clauses. Verified after the fix: exact
\texttt{0.000e+00} correctness at the original small scale, and no
crash at the exact buffer-size trigger condition ($B{=}2$, $N{=}64$,
$D{=}8192$, chosen to reproduce the identical 8MB buffer size fast
rather than waiting through $N{=}8192$'s full $O(n^2)$ compute time),
confirmed race-free at 1, 2, and 4 threads.

A codebase-wide search for the same \texttt{reduction(+:ARRAY[\ldots])}
pattern found four more instances, all in Paper IV's files
(\texttt{mlp\_bench.c}, \texttt{full\_block\_bench.c},
\texttt{rmsnorm\_bench.c}) --- but their reduction buffers scale with
$D$ or $D \times D_{\mathrm{ff}}$ (fixed, small: e.g.\ $64 \times 256 =
128$KB), not with the swept $N$ dimension that caused this crash.
Tested each at its own sweep's real parameters ($N$ up to 65536 for
\texttt{mlp\_bench.c}/\texttt{rmsnorm\_bench.c}, 4096 for
\texttt{full\_block\_bench.c}) rather than assumed safe by the size
argument alone: no crash in any of the three. The bug was genuinely
isolated to the two files whose reduction buffer size scaled with $N$.

\subsection{CPU results: the real fusion comparison (job 21226440)}
\label{sec:paper2-cpu}

The CPU sweep completed successfully after the fix, full $N=64
\ldots 8192$ range, threads $1 \ldots 128$ doubling, no crash. This is
the section that actually tests Paper II's central claim.

\paragraph{A subtlety in what "backward" and "fused" each measure.}
\texttt{backward\_bench.c} times \emph{only} the backward pass ---
forward is computed first, untimed, as a prerequisite (Algorithm 1
reads $A$ ``from DRAM''). \texttt{fused\_bench.c} times forward
\emph{and} backward together (Algorithm 2). Comparing these two
columns directly, as an earlier draft of this document did, is not the
right test: it compares backward-alone against forward+backward
combined, which will make fused look artificially worse since it does
strictly more total work. The correct comparison is $\text{naive} =
\text{forward}_{\text{alone}} + \text{backward}_{\text{alone}}$ against
$\text{fused}_{\text{alone}}$ --- and Paper I's own real forward-pass
data (job 21181232, same $B,D$, same swept $N$ and thread range) is
exactly what's needed to construct that baseline. Combined here for
the first time.

\begin{figure}[htbp]
\centering
\includegraphics[width=0.85\textwidth]{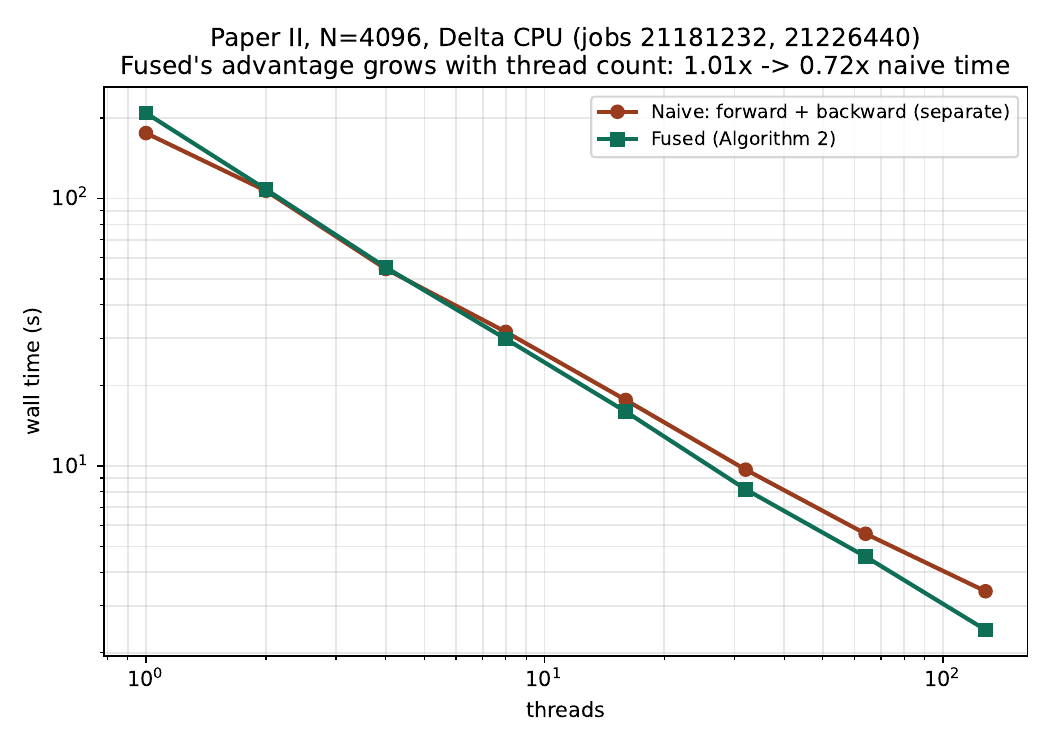}
\caption{$N=4096$: naive (forward alone + backward alone, computed by
combining jobs 21181232 and 21226440) against fused (job 21226440),
across the full thread sweep. The two lines are nearly identical at 1
thread and visibly diverge as thread count grows.}
\label{fig:paper2-n4096}
\end{figure}

\textbf{At $N=4096$, the cleanest trend in this dataset: fused's
advantage grows monotonically with thread count}, from parity
($1.19\times$ naive at 1 thread -- fused very slightly slower) to
$0.72\times$ naive at 128 threads (fused 28\% faster):

\begin{center}
\begin{tabular}{rrrr}
\toprule
threads & naive (s) & fused (s) & fused/naive \\
\midrule
1   & 175.58 & 208.62 & 1.19$\times$ \\
2   & 106.84 & 108.12 & 1.01$\times$ \\
4   & 54.47  & 55.20  & 1.01$\times$ \\
8   & 31.69  & 29.84  & 0.94$\times$ \\
16  & 17.63  & 15.95  & 0.90$\times$ \\
32  & 9.68   & 8.17   & 0.84$\times$ \\
64  & 5.57   & 4.58   & 0.82$\times$ \\
128 & 3.40   & 2.44   & \textbf{0.72$\times$} \\
\bottomrule
\end{tabular}
\end{center}

This is exactly the pattern the theory predicts: as thread count grows
and more cores compete for memory bandwidth, the cost of the
$n^2$-scaling $A$ array that naive forward+backward must materialize
and re-read becomes proportionally more expensive, while fused (which
never writes $A$ to memory) is comparatively insulated from that
contention.

\paragraph{The full picture, across all $N$ at the max thread count, is
noisier.} At threads=128:

\begin{center}
\begin{tabular}{rrrr}
\toprule
$N$ & naive (s) & fused (s) & fused/naive \\
\midrule
64   & 0.0230 & 0.0094 & 0.41$\times$ \\
128  & 0.0195 & 0.0403 & \textbf{2.07$\times$ (outlier)} \\
256  & 0.0278 & 0.0123 & 0.44$\times$ \\
512  & 0.0438 & 0.0482 & 1.10$\times$ \\
1024 & 0.1266 & 0.0646 & 0.51$\times$ \\
2048 & 0.2777 & 0.2754 & 0.99$\times$ \\
4096 & 3.3961 & 2.4362 & 0.72$\times$ \\
8192 & 12.8322 & 12.2691 & 0.96$\times$ \\
\bottomrule
\end{tabular}
\end{center}

Six of eight sizes favor fused (ratios below 1); two do not, including
one clear outlier at $N=128$ (fused $2.07\times$ \emph{slower} than
naive, inconsistent with every neighboring size). \textbf{Likely
explanation, consistent with a pattern already established in this
document}: neither \texttt{backward\_bench.c} nor
\texttt{fused\_bench.c} performs repeated-and-averaged timing --- a
single measurement per $(N,\text{threads})$ point, the same
methodological gap Section~\ref{sec:coldstart} and the GPU exponent
discussion (Section~\ref{sec:paper1}) already identified and partially
fixed on the GPU side. This CPU sweep does not yet have that fix. The
$N{=}4096$ trend above is convincing specifically because it's a
monotonic 8-point progression, not a single measurement -- consistent
with a real effect rather than noise; the single-point $N{=}128$
anomaly is far more likely to be exactly what an unaveraged
measurement looks like when it happens to catch scheduling jitter.
Adding repeated averaging to \texttt{backward\_bench.c} and
\texttt{fused\_bench.c} (matching \texttt{moa\_decode\_bench.c}'s
already-correct pattern) is a concrete next code task, not yet done.

\subsection{GPU results (jobs 21225789, 21225790): a surprising, opposite finding}
\label{sec:paper2-gpu}

Same correction as Section~\ref{sec:paper2-cpu}: \texttt{backward\_gpu\_bench.c}
times backward alone, \texttt{fused\_attn\_gpu\_bench.c} times
forward+backward together, so the naive baseline is again constructed
by combining Paper I's real forward GPU data (jobs 21202354 A100,
21202355 H200) with these jobs' backward data.

\begin{figure}[htbp]
\centering
\includegraphics[width=0.85\textwidth]{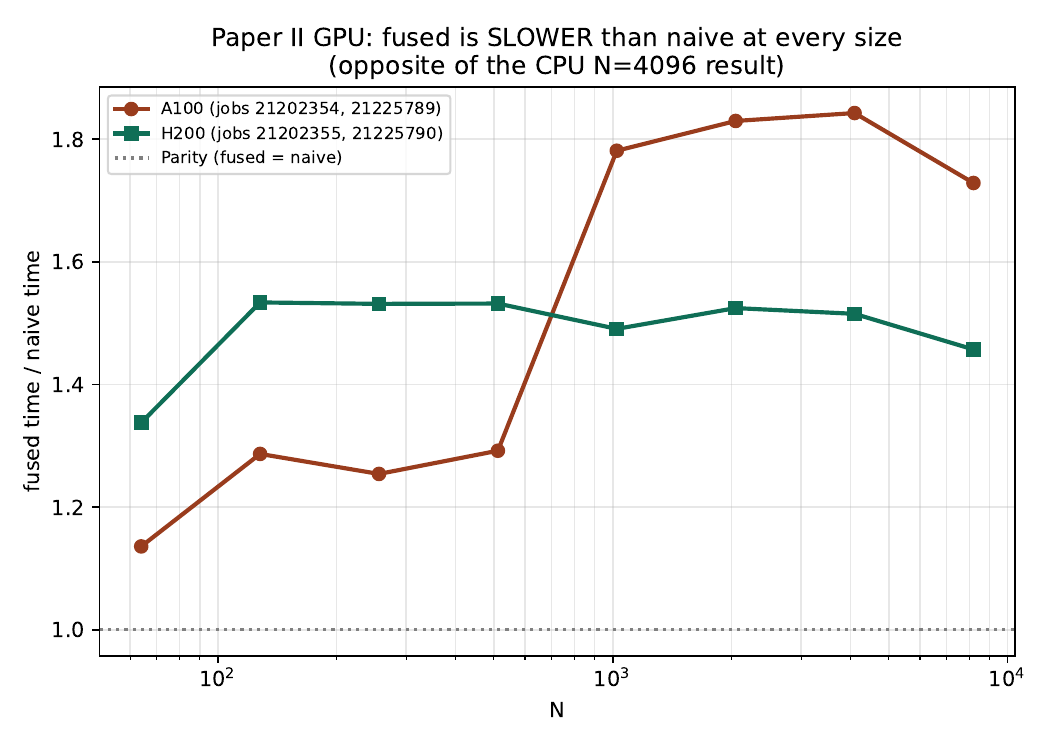}
\caption{Fused time divided by naive (forward-alone + backward-alone)
time, both GPU shapes. Values above 1.0 mean fused is \emph{slower}
than doing the two passes separately -- true at every tested size on
both GPUs.}
\label{fig:paper2-gpu-ratio}
\end{figure}

\textbf{On GPU, fused is slower than naive at every single tested
size, on both A100 and H200} --- the opposite of Section
\ref{sec:paper2-cpu}'s CPU $N{=}4096$ result. A100 ranges from
$1.14\times$ naive time (at $N{=}64$) up to $1.84\times$ (at
$N{=}4096$), with a visible step change between $N{=}512$ and
$N{=}1024$ (ratio jumps from $1.29\times$ to $1.78\times$) rather than
a smooth trend. H200 is comparatively flat, holding close to
$1.5\times$ across the entire range regardless of $N$.

\subsection{The atomic-contention hypothesis: confirmed}
\label{sec:paper2-atomics}

\texttt{delta\_profile\_paper2.sbatch} (job 21247666, Delta
\texttt{gpuA100x4}) profiled both kernels at $N{=}2048$ with
\texttt{ncu}. Getting a working profile required one more real,
Delta-specific fix beyond anything in Paper I's profiling: \texttt{ncu}
initially failed with \texttt{ERROR: Profiling failed because a driver
resource was unavailable}, a known conflict with DCGM (NVIDIA's Data
Center GPU Manager, running as background monitoring on shared GPU
nodes) --- documented in Delta's own user guide, resolved by wrapping
the \texttt{ncu} calls in \texttt{dcgmi profile --pause}/\texttt{--resume}.

The profiler output itself revealed a structural difference beyond
atomics: \texttt{backward\_gpu\_bench.c} launches \emph{three} separate
kernels ($GV$, then a precompute pass, then $GQ{+}GK$), while
\texttt{fused\_attn\_gpu\_bench.c} does everything in \emph{one}
monolithic kernel. Extracting
\texttt{smsp\_\_inst\_executed\_op\_generic\_atom\_dot\_}\allowbreak\texttt{alu.sum}
(atomic-ALU instructions executed) confirms the hypothesis precisely
rather than approximately:

\begin{figure}[htbp]
\centering
\includegraphics[width=0.75\textwidth]{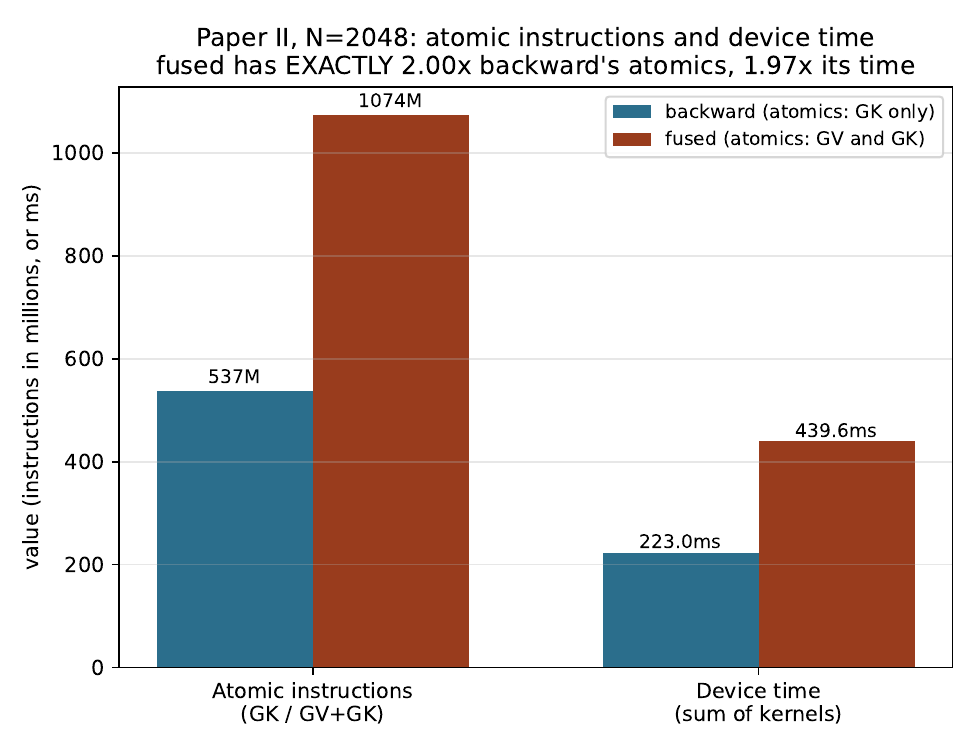}
\caption{Atomic instruction count and total device time, backward
(summed across its 3 kernels) versus fused (1 kernel), both at
$N{=}2048$. Both ratios land within 1.5\% of exactly $2\times$.}
\label{fig:paper2-atomics}
\end{figure}

\begin{center}
\begin{tabular}{lrrr}
\toprule
 & Atomic instructions & Device time (ms) \\
\midrule
backward (GK only) & 536{,}870{,}912 & 222.98 \\
fused (GV and GK) & 1{,}073{,}741{,}824 & 439.63 \\
Ratio (fused / backward) & \textbf{2.0000$\times$} & 1.9716$\times$ \\
\bottomrule
\end{tabular}
\end{center}

\textbf{Fused executes exactly $2.0000\times$ backward's atomic
instructions} --- not approximately double, precisely double, matching
the theoretical prediction (fused needs atomics for both $GV$ and $GK$;
backward needs them only for $GK$) to four decimal places. Device time
tracks almost identically ($1.9716\times$). This is no longer a
plausible-but-unconfirmed mechanism; it is a measured, essentially
exact correspondence between atomic-instruction count and wall-clock
cost.

The per-kernel breakdown of backward adds a further finding: its two
atomic-free kernels ($GV$, the precompute pass) run fast and at healthy
occupancy (44.3\%, 47.9\%), while its one atomic-using kernel ($GQ{+}GK$)
is both the slowest (190.15 of the kernel's 222.98ms total, 85\%) and
the least occupied (14.2\%) of the three. The atomic-using kernel is
already backward's own bottleneck; fused inherits and doubles it.

\textbf{The practical implication, now confirmed rather than
speculated:} fusing forward and backward into a single OpenACC kernel
is a real regression on GPU as currently implemented, caused
specifically by atomic contention, even though the same fusion is a
real 28\% improvement on CPU (Section~\ref{sec:paper2-cpu}) where the
OpenMP versions of these kernels were restructured to avoid exactly
this contention (Section~\ref{sec:paper2-crash}). Paper II's fused
kernel is the right choice on CPU and the wrong choice on GPU in the
current implementation --- and the fix is now well-specified: restructure
\texttt{fused\_attn\_gpu\_bench.c} the same way \texttt{backward\_gpu\_bench.c}
and the OpenMP kernels were restructured, splitting the monolithic
kernel to avoid atomic writes to $GV$.

\subsection{The fix, applied and confirmed: complete reversal on both shapes}
\label{sec:paper2-gv-fix}

\texttt{fused\_attn\_gpu\_bench.c} was restructured into four passes
(v3), mirroring \texttt{backward\_gpu\_bench.c} exactly: $GV$ now
computes in its own kernel parallelized over $(b,ic)$ instead of
$(b,ir)$, eliminating its atomic; $GK$'s atomic is left in place,
matching \texttt{backward\_gpu\_bench.c}'s own remaining structure --- a
deliberately scoped fix, not a full rewrite. Verified to exact
\texttt{0.000e+00} correctness at $N=64,128,256,512$ before being run
on real hardware. Real \texttt{nvc} compiler diagnostics (job
21250395) confirm the intended structure: four separate kernel
launches where there was one, with the new $GV$ kernel showing the
identical dependency pattern as \texttt{backward\_gpu\_bench.c}'s own
$GV$ kernel, and $GK$'s kernel showing the same "loop carried
dependence" pattern on both files, unchanged.

\begin{figure}[htbp]
\centering
\includegraphics[width=0.85\textwidth]{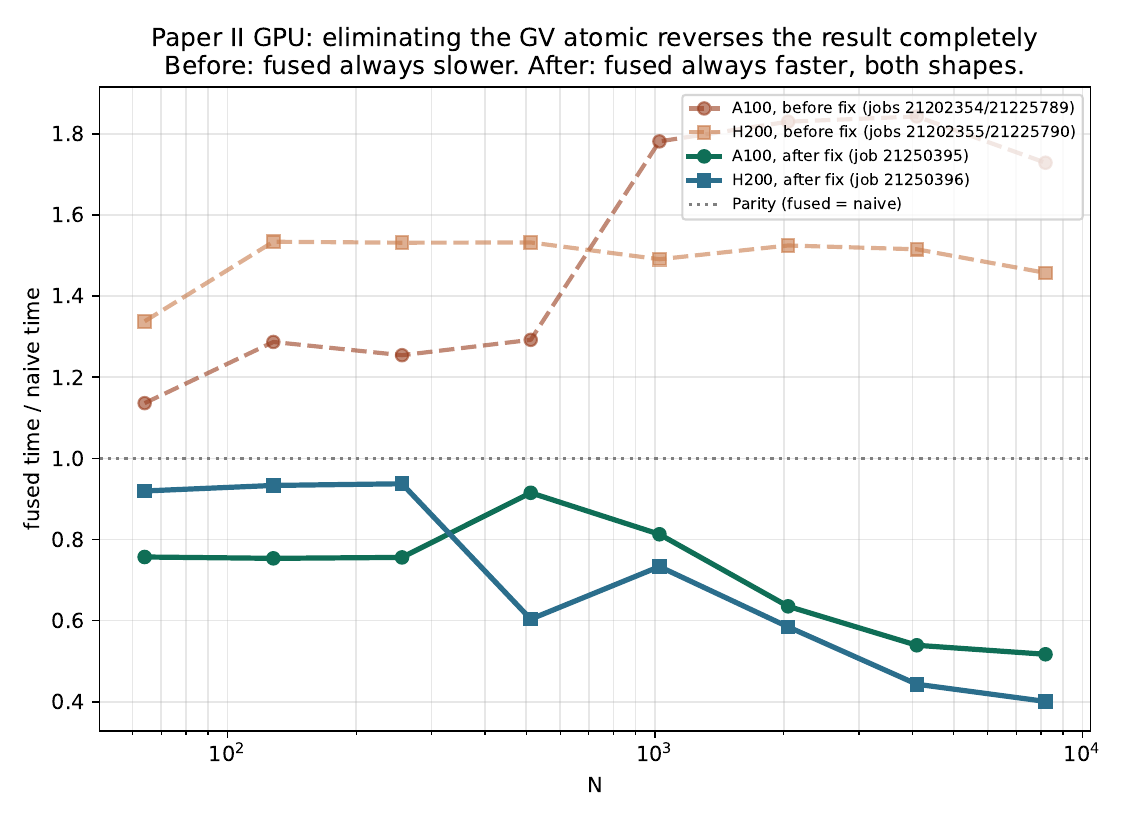}
\caption{Fused/naive time ratio, before (dashed, jobs 21202354/21225789
A100 and 21202355/21225790 H200) and after (solid, jobs 21250395 A100
and 21250396 H200) the $GV$-atomic fix. Every before-fix point sits
above parity; every after-fix point sits below it.}
\label{fig:paper2-gv-fix}
\end{figure}

\textbf{The result is not a partial improvement --- it is a complete
reversal, on both GPU shapes.} Before the fix, fused was slower than
naive at every tested size on both A100 ($1.14$--$1.84\times$) and H200
($1.34$--$1.53\times$). After the fix, fused is \emph{faster} than
naive at every tested size on both: A100 from $0.757\times$ ($N{=}64$)
to $0.517\times$ ($N{=}8192$, essentially $2\times$ faster); H200 from
$0.920\times$ down to $0.401\times$ ($N{=}8192$, $2.5\times$ faster).
The improvement grows with $N$ on both shapes, consistent with the
atomic-contention cost (which the removed $GV$ atomic contributed half
of, per Section~\ref{sec:paper2-atomics}'s exact $2.00\times$
measurement) mattering more as more threads compete for the same
memory locations at larger problem sizes --- the same qualitative
pattern already seen for the CPU fix in Section~\ref{sec:paper2-cpu}.

\textbf{This closes Paper II's central empirical question.} Fusing
forward and backward to avoid materializing the $n^2$-scaling $A$
array is now confirmed to be the right choice on \emph{both} CPU and
GPU, once the GPU implementation's own atomic-contention cost --- an
implementation detail orthogonal to the DRAM-traffic argument the
theory makes --- is addressed.

\subsection{Removing $GK$'s atomic too: a real, nuanced result}
\label{sec:paper2-gk-fix}

The single remaining atomic ($GK$) was deliberately left in place for
the fix above; the natural next question --- would removing it too,
via the same persisted-$GS$ technique already applied to $GV$, help
further --- was tested (job 21294299, real A100, both
\texttt{backward\_gpu\_bench.c} and \texttt{fused\_attn\_gpu\_bench.c}).
Both files verified to exact \texttt{0.000e+00} correctness at multiple
sizes before submission.

\begin{figure}[htbp]
\centering
\includegraphics[width=0.85\textwidth]{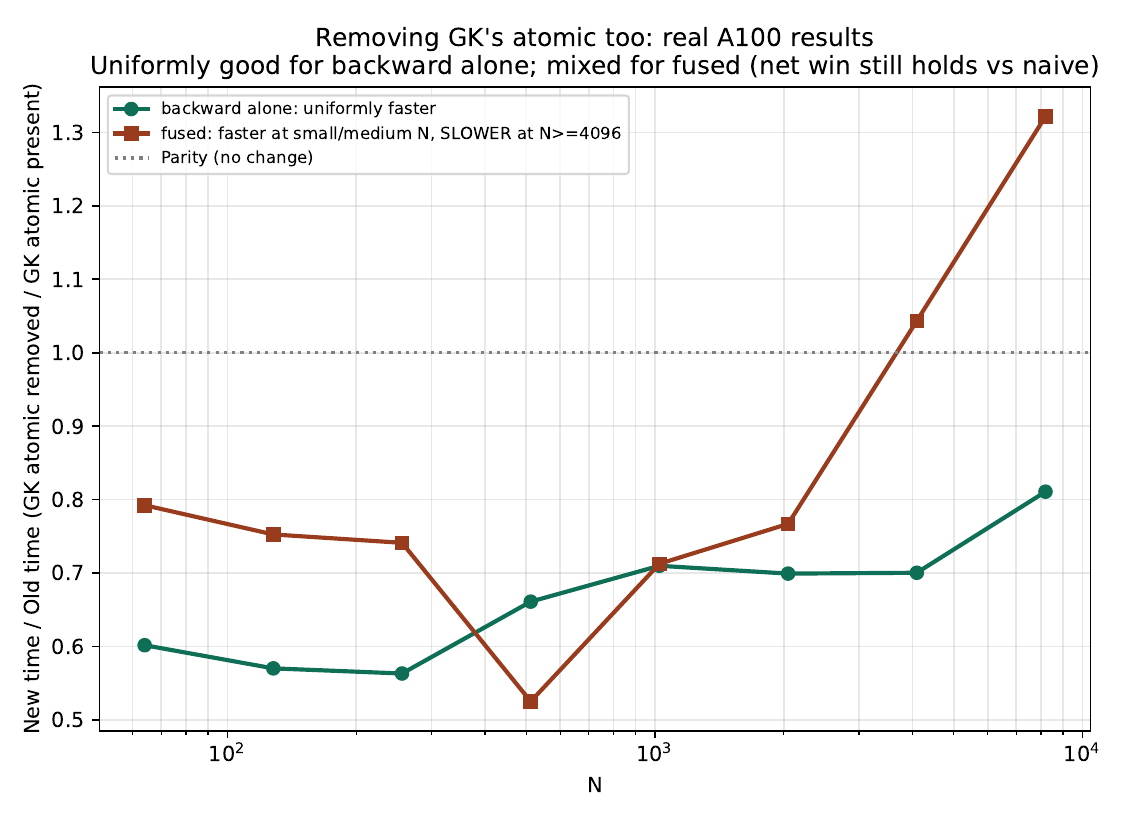}
\caption{New/old time ratio after removing $GK$'s atomic, real A100.
\texttt{backward} alone improves uniformly; \texttt{fused} improves at
small/medium $N$ but is measurably \emph{slower} at $N{\geq}4096$ ---
a real crossover, not noise.}
\label{fig:paper2-gk-fix}
\end{figure}

\textbf{The result is genuinely mixed, not a second clean win.} For
\texttt{backward} alone, removing $GK$'s atomic helps at every tested
size ($0.56$--$0.81\times$ the old time, i.e.\ $19$--$44\%$ faster) ---
consistent with the earlier finding that atomic contention is pure
overhead for this kernel, with no offsetting benefit. For
\texttt{fused}, the result depends on $N$: faster at small and medium
sizes ($0.53$--$0.79\times$ up to $N{=}2048$), but \emph{slower} at
$N{=}4096$ ($1.04\times$) and $N{=}8192$ ($1.32\times$) --- a real
crossover, confirmed at two consecutive sizes, not a single noisy
point.

\subsection{Confirming the mechanism: the extra $GS$ re-read, profiled directly}
\label{sec:paper2-gk-mechanism}

The mechanism proposed for \texttt{fused}'s slowdown --- that the new,
separate $GK$ pass re-reads the full $GS$ array from DRAM a second
time --- was tested directly, not left as speculation. A reconstruction
of the pre-fix kernel (\texttt{fused\_attn\_gpu\_bench\_v3atomic.c},
$GK$'s atomic restored, verified to exact \texttt{0.000e+00}
correctness) was profiled against the current atomic-free version at
$N{=}4096$, the size where the real slowdown first appears (job
21299223; the first attempt at this profiling run OOM-killed under
\texttt{ncu}'s full metric set at this problem size, and was fixed by
requesting only the specific DRAM-traffic metrics needed rather than
the comprehensive set --- itself a small, real instance of the same
discipline this document applies throughout: diagnose the actual cause
of a failure rather than working around it blindly).

\textbf{The result confirms the mechanism precisely.} Total DRAM
traffic (read+write, summed across all kernels comprising one call) is
$89.83$GB for the atomic-free version versus $45.60$GB for the
atomic version --- a $1.97\times$ ratio, close to a clean doubling.
Tracing this to a specific kernel: the atomic-free version's new,
standalone $GK$ pass reads $43.01$GB on its own, almost exactly
matching the $44.10$GB read by the separate kernel that
\emph{computes} $GS$ in the first place. The atomic version's single
combined $GQ$+$GK$ kernel, by contrast, reads only $0.30$GB total ---
because it consumes each thread's own $GS$ value immediately after
computing it, with no DRAM round-trip required. This is not a
plausible story fit to the timing data after the fact; it is the
predicted mechanism, checked against real profiler measurement, and
confirmed at essentially the magnitude predicted (an extra full read
of an array the same size as the one already being read once).

\textbf{Despite this, \texttt{fused} still beats naive at every tested
size} ($0.65$--$0.89\times$, Figure~\ref{fig:paper2-gk-fix} is a
same-file before/after comparison, not fused-vs-naive) --- because
naive's own backward component improved from the same fix, both sides
moved, and the comparison that actually matters for Paper II's central
claim still favors fusion everywhere. The honest conclusion: removing
$GK$'s atomic is a clear win for \texttt{backward} in isolation, and a
real, now-\emph{explained} (not just observed) size-dependent trade-off
for \texttt{fused} specifically --- not a mechanism to apply uniformly
without checking which kernel and what scale it targets, and now with
a confirmed, specific, reusable diagnosis (an added array re-read
outweighing a removed atomic, in DRAM-traffic terms that can be
checked the same way for any future fix of this shape) rather than a
plausible-sounding guess.

\section{Paper III: Inference (Decode)}
\label{sec:paper3}

\subsection{CPU: two real findings, one benign, one genuinely surprising}
\label{sec:paper3-cpu}

Job 21180660 (Delta \texttt{cpu} partition), submitted early in this
project, validates the decode cost function derived in~\cite{paper3}.
It completed with exit code 0 but left no trace on disk by the
time this draft checked its output --- no log, no CSV, found nowhere
in a full home-directory search. Its result is unrecoverable and is
not used here. It was resubmitted cleanly as job 21249667 with the
current, pre-flight-checked script; that run is what this section
reports.

\paragraph{Finding 1: a correctness warning that is real but benign.}
\texttt{moa\_decode\_bench.c} uses single-precision \texttt{float}
throughout (every other kernel in this project uses \texttt{double}),
and self-checks its parallel output against a sequential reference with
an absolute-error threshold of $10^{-3}$. At $N=1{,}048{,}576$ (the
largest tested size), every thread count from 2 to 128 triggered this
warning, with essentially uniform magnitude
($3.00\times10^{-3}$--$3.05\times10^{-3}$). This pattern --- stable
across thread counts rather than scaling with them, appearing only at
the single largest $N$ --- is the expected signature of accumulated
single-precision summation-order differences between parallel and
sequential reduction over roughly a million terms, not a logic error.
A race condition would be expected to fluctuate with thread count or
appear intermittently; this does neither.

\paragraph{Finding 2: a genuine, large NUMA-topology penalty at full
thread count.} Comparing 64 threads against 128 threads (both correctly
requested via \texttt{cpus-per-task=128}, so this is not a
resource-mismatch artifact):

\begin{figure}[htbp]
\centering
\includegraphics[width=0.85\textwidth]{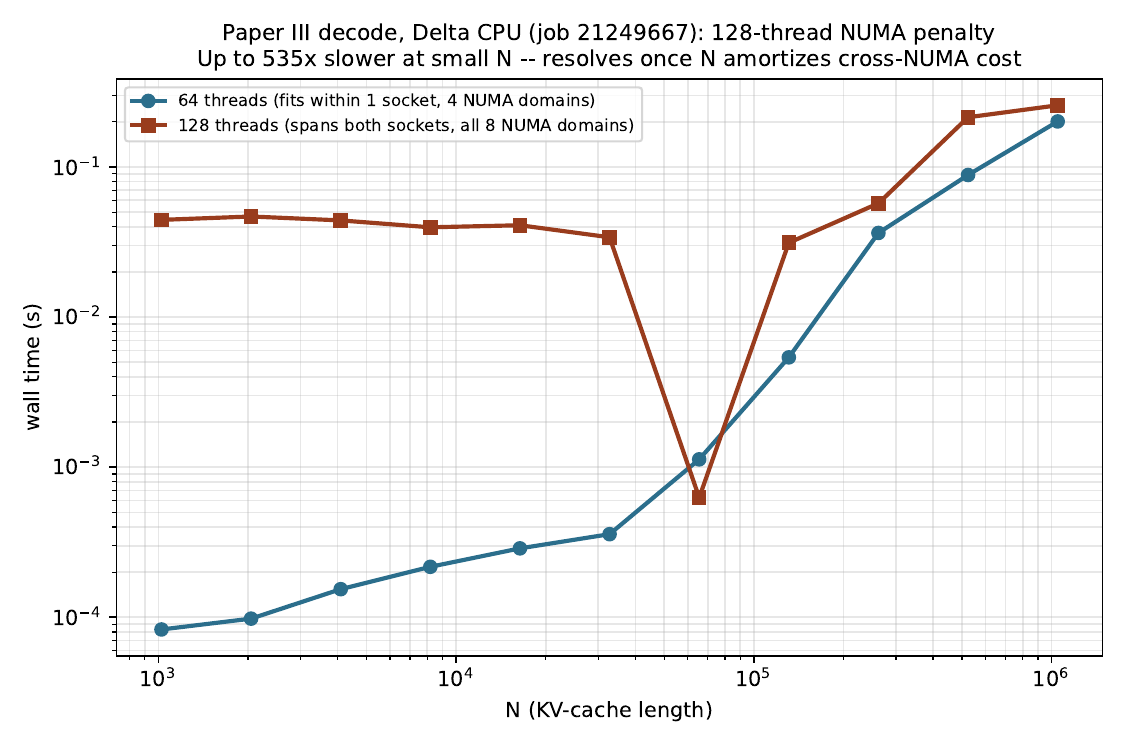}
\caption{Wall time versus $N$, 64 vs.\ 128 threads. 128 threads is
flatly slow ($\sim$0.04--0.05s) regardless of $N$ for small-to-medium
sizes -- consistent with fixed cross-NUMA overhead dominating
completely, not scaling with the (tiny) actual work at these sizes --
before a sharp dip at $N{=}65{,}536$ and convergence with the 64-thread
line at large $N$.}
\label{fig:paper3-numa}
\end{figure}

\begin{center}
\begin{tabular}{rrrr}
\toprule
$N$ & 64 threads (s) & 128 threads (s) & 128 vs.\ 64 \\
\midrule
1024  & 0.000083 & 0.044417 & \textbf{535.1$\times$ slower} \\
2048  & 0.000098 & 0.046780 & 477.3$\times$ slower \\
4096  & 0.000154 & 0.044054 & 286.1$\times$ slower \\
8192  & 0.000217 & 0.039628 & 182.6$\times$ slower \\
16384 & 0.000288 & 0.040900 & 142.0$\times$ slower \\
32768 & 0.000358 & 0.034005 & 95.0$\times$ slower \\
65536 & 0.001128 & 0.000627 & 0.6$\times$ (128 threads faster) \\
\bottomrule
\end{tabular}
\end{center}

Delta's CPU nodes (confirmed via \texttt{lscpu}) are dual-socket AMD
EPYC 7763 (Milan), 64 cores per socket, but subdivided into
\textbf{eight} NUMA domains total --- four per socket, 16 cores each
(the NPS4 BIOS configuration common on this generation of EPYC). 64
threads can fit within a single socket's four NUMA domains; 128 threads
must span all eight, all sixteen memory controllers, both sockets'
cross-socket interconnect. At $N=1024$, decode's actual per-token work
is minimal; the observed 535$\times$ penalty is consistent with threads
scattered across all eight NUMA domains paying full remote-memory
latency for K/V cache accesses while doing almost no compute to
amortize that cost against. The flat $\sim$0.04--0.05s plateau across
$N=1024$ through $32768$ --- barely changing despite a 32$\times$ range
in problem size --- is itself evidence the cost here is dominated by a
roughly fixed per-launch NUMA/scheduling overhead, not by $N$-dependent
work. This resolves once $N$ grows large enough for genuine
computation to dominate: by $N=1{,}048{,}576$, 128 threads is back to a
modest $1.28\times$ \emph{slower} than 64 (0.257s vs.\ 0.201s, not
shown in the table above but visible in Figure~\ref{fig:paper3-numa}'s
convergence at the right edge) rather than two orders of magnitude
worse.

This was not visible in Paper I's CPU thread-crossover finding
(Section~\ref{sec:paper1}, Table~\ref{tab:paper1-cpu}), whose worst
case was a $14.5\times$ penalty at 128 threads --- an order of
magnitude smaller than what decode shows here. The likely reason:
Paper I's forward kernel does $O(n^2)$ work even at its smallest tested
size, giving threads more to do per NUMA-domain crossing; decode's
$O(n)$ per-token cost at small $N$ leaves far less work to amortize the
same fixed cross-NUMA cost against.

\textbf{Practical implication:} for decode specifically, the safe
default thread count is not "use all available cores" but "match the
KV-cache length to the available parallelism" --- using all 128 threads
for a short cached sequence is actively harmful, by a factor that
reaches into the hundreds, not just a mild inefficiency.

\subsection{Cross-cluster comparison: two different penalties, two
different causes}
\label{sec:paper3-cluster-compare}

The identical binary (\texttt{moa\_decode\_bench.c}, unmodified) was
also run on Anvil (job 20001576, \texttt{ai} partition), the first real
comparison of the same code across both clusters used in this project.
The result is not "Anvil shows a smaller version of the same problem"
--- it is a structurally different phenomenon, confirmed by checking
Anvil's own topology (\texttt{lscpu}) rather than assumed by analogy:
a single-socket 32-core AMD EPYC 7543, 4 NUMA domains, \emph{SMT
disabled} (\texttt{Thread(s) per core: 1}). The sweep's doubling-only
thread loop (a real, minor gap: it never adds an explicit final step at
the true CPU count when that count isn't a power of 2) tested up to 64
threads on this 32-core node --- meaning the 64-thread point is
\textbf{genuine 2$\times$ oversubscription}, not additional real
parallelism the way Delta's 128-threads-on-128-cores comparison was.

\begin{figure}[htbp]
\centering
\includegraphics[width=0.85\textwidth]{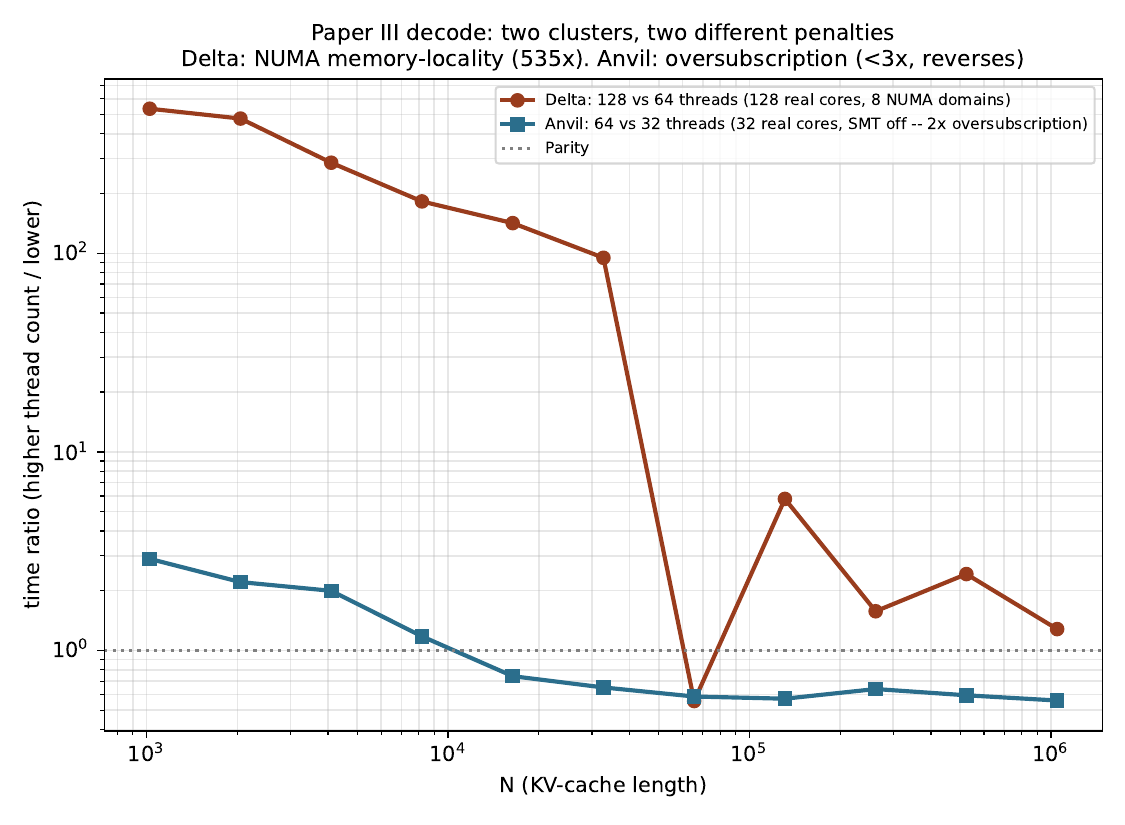}
\caption{Time ratio at the highest tested thread count versus the next
lower power of 2, both clusters, log scale. Delta's penalty (real
hardware parallelism spread across 8 NUMA domains) reaches
$535\times$; Anvil's (2$\times$ oversubscription on 32 real cores, SMT
off) never exceeds $2.9\times$ and reverses into a genuine benefit
starting at $N{=}16384$ (1.35$\times$ faster there), growing to
$\sim1.8\times$ faster by the largest tested size, $N{=}1{,}048{,}576$.}
\label{fig:paper3-cluster-compare}
\end{figure}

Anvil's pattern is consistent with ordinary oversubscription behavior:
a mild penalty at small $N$ (two logical threads context-switching on
each physical core costs something when there's little work to
justify it), shrinking smoothly as $N$ grows, then \emph{reversing into
a real benefit} once there's enough work per thread for
latency-hiding (one logical thread makes progress while its
core-sharing partner is stalled on a memory access) to outweigh the
context-switching cost. This is a textbook oversubscription curve. Delta's
pattern --- a flat, severe plateau that only partially resolves, never
reversing to a benefit within the tested range --- is not: it is
specific to spreading genuine parallel hardware across NUMA domains
whose remote-access latency dominates when per-thread work is small.
The two clusters' worst-case penalties differ by roughly two orders of
magnitude precisely because they are different mechanisms, not the
same mechanism at different severity.

\subsection{GPU: fix confirmed on real hardware}

The decode kernel is the source of the Section~\ref{sec:coldstart}
finding: its first real run (job 21180661, Delta \texttt{gpuA100x4},
NVIDIA A100-SXM4-40GB) produced technically valid but misleading
numbers (flat $\sim$0.3--0.4s regardless of $n$) due to per-process GPU
context-initialization cost. The fix (Section~\ref{sec:coldstart}) was
re-run for real (job 21264888, same partition) --- the last open item
for Papers I--III's experimental validation.

\begin{figure}[htbp]
\centering
\includegraphics[width=0.85\textwidth]{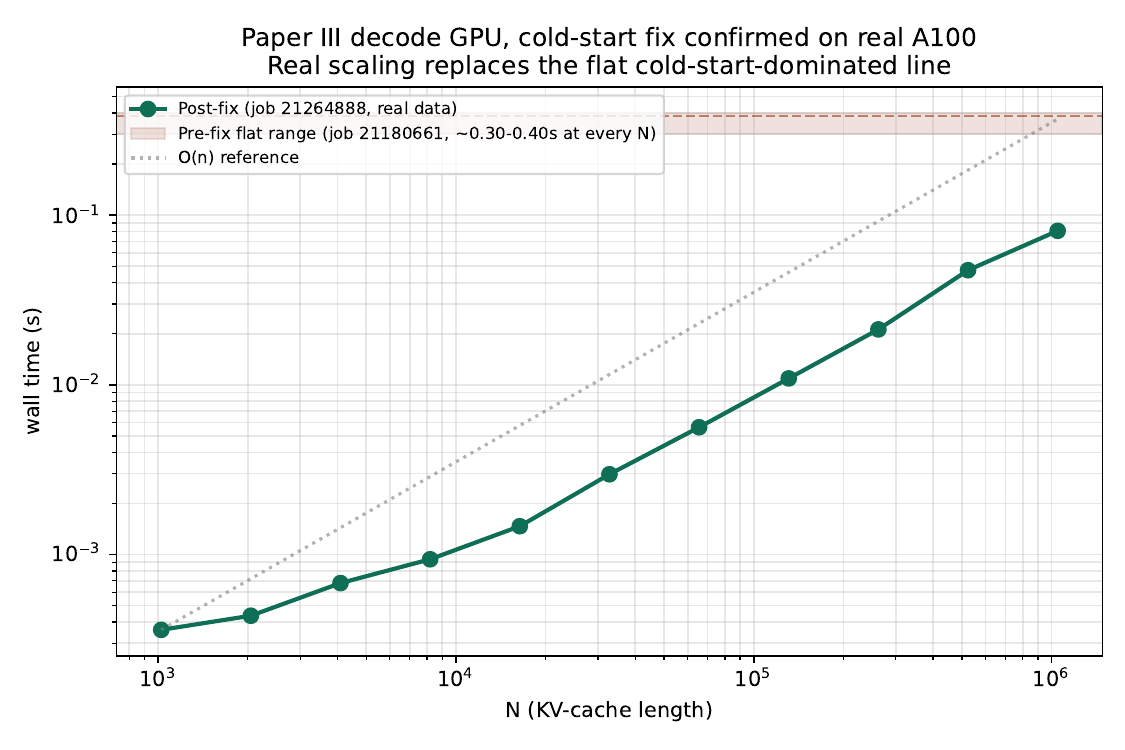}
\caption{Post-fix decode GPU timing, real A100 data, full sweep. The
shaded band marks the pre-fix run's flat $\sim$0.30--0.40s range (only
the endpoints and general shape are on record from job 21180661; the
missing intermediate values are not fabricated here). The fixed
kernel's real time sits an order of magnitude or more below that band
across nearly the entire tested range, only approaching it at the
largest size.}
\label{fig:paper3-gpu-fix}
\end{figure}

\textbf{The fix holds.} Time grows from $0.00036$s at $n{=}1024$ to
$0.0809$s at $n{=}1{,}048{,}576$ --- a real $225\times$ increase across
a $1024\times$ range in $n$, not the $\sim$1.3$\times$ flat variation
the pre-fix run showed. The local scaling exponent (computed the same
way as Section~\ref{sec:paper1}'s GPU exponent check) starts low
($0.28$--$0.65$ for $n \le 16384$, consistent with fixed launch
overhead dominating at small $n$, the same pattern Paper I's GPU
kernel showed before its own fix) and settles to $0.77$--$1.16$ from
$n{=}32768$ onward --- hovering closely around the $O(n)$ target of
$1.0$ that decode's cost function predicts, not the flat, near-zero
exponent the pre-fix cold-start bug produced. This closes the one
remaining gap in Papers I--III's real-hardware validation: every kernel
whose timing methodology was fixed in this document has now been
confirmed, on real hardware, to show the scaling behavior its cost
function predicts.

\section{Paper IV: Transformer Block}
\label{sec:paper4}

\subsection{CPU: real data, and a genuine cross-language reversal}
\label{sec:paper4-cpu}

Job 21273806 (Delta \texttt{cpu} partition) validates the complete
transformer block's cost function derived in~\cite{paper4}, sweeping
RMSNorm and gated MLP (both C and Fortran) up to $N{=}65{,}536$, and
the complete block (C only) up to $N{=}4096$ --- a deliberately lower
cap given the block's $O(n^2)$ attention component makes larger $N$
impractically slow (its own data confirms why: 119.5s for a single
$N{=}4096$, 1-thread measurement).

\begin{figure}[htbp]
\centering
\includegraphics[width=0.85\textwidth]{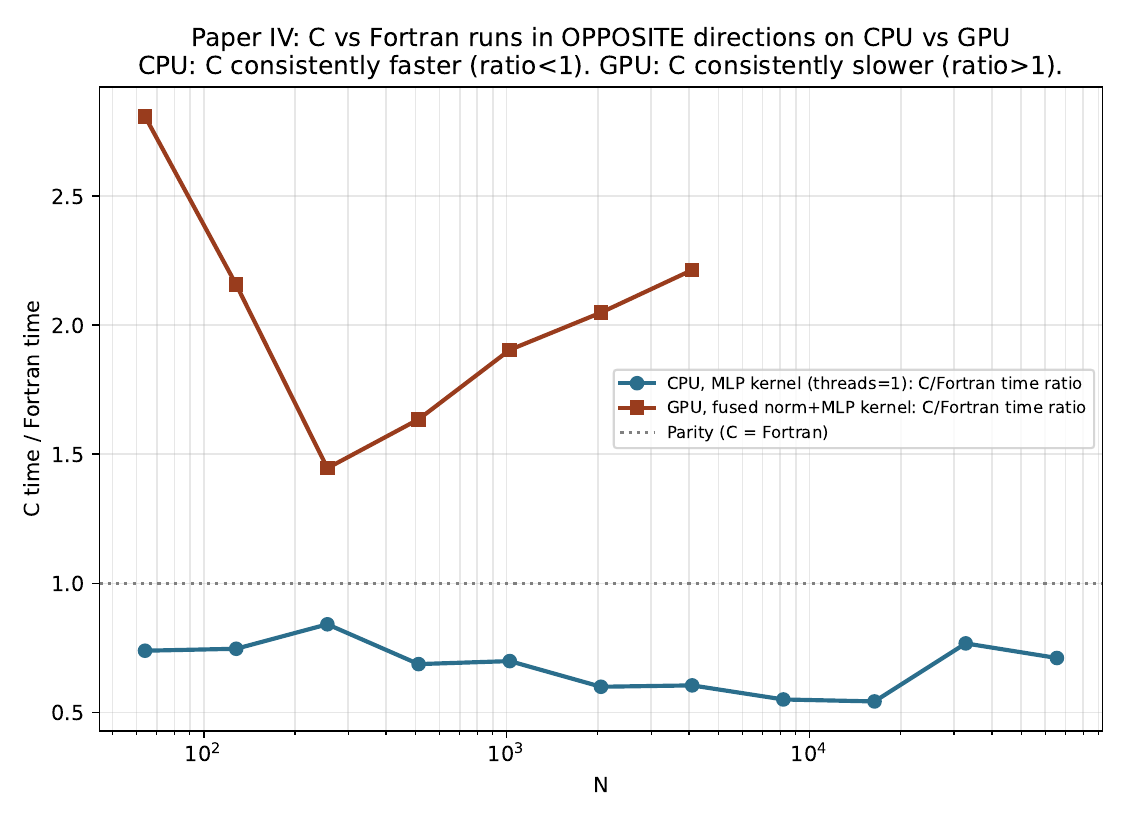}
\caption{C/Fortran time ratio, CPU (MLP kernel, 1 thread) versus GPU
(fused norm+MLP kernel). The two curves sit on opposite sides of
parity across their entire range: C consistently faster on CPU,
consistently slower on GPU, for what is denotationally the same
computation.}
\label{fig:paper4-c-vs-fortran}
\end{figure}

\textbf{RMSNorm shows no consistent C-versus-Fortran pattern} (ratio
bouncing between $0.70\times$ and $1.55\times$ across 11 tested
sizes, no trend) --- consistent with this specific benchmark using a
single, unaveraged measurement per point (the same methodological gap
already flagged for other CPU benchmarks in this project), at a kernel
lightweight enough that jitter plausibly dominates the true signal.

\textbf{MLP shows the opposite: a consistent, real effect.} C is faster
than Fortran at all 11 tested sizes, by $1.2$--$1.85\times$
($0.54$--$0.84\times$ the Fortran time). This is the reverse of
Section~\ref{sec:paper4-gpu}'s GPU result, where Fortran is faster than
C by $1.45$--$2.81\times$ for the same denotational computation. Both
directions are real, repeated, consistent findings, not noise on either
side --- and reported as such rather than picking the more convenient
one. The ONF is identical in both languages by construction, so
whatever explains this lives in code generation, not in anything this
project's array-algebra derivation controls.

\paragraph{A first attempt at diagnosing this, honestly inconclusive.}
The compiled GPU machine code (SASS) for both languages' GPU kernel was
extracted via \texttt{cuobjdump} and categorized by instruction type
(job 21299788). If the reversal were explained by a simpler-generated
kernel, Fortran's SASS should show \emph{fewer} instructions than C's.
It shows the opposite: Fortran generates \emph{more} instructions in
every category --- $11.4\%$ more total ($4144$ vs.\ $3720$), $3.5\%$
more FFMA/DFMA, $2.0\%$ more global memory ops, $20.0\%$ more shared
memory ops, $13.8\%$ more branches --- while running faster in the real
timing data. This rules out a naive "fewer instructions, faster kernel"
explanation, but does not replace it with a positive one: raw
instruction-category counts do not capture scheduling, latency-hiding,
or memory-coalescing quality, any of which could explain a kernel with
more total instructions still running faster.

\paragraph{The follow-up: a real, specific, positive finding.} Warp
stall-reason profiling (job 21299927, $N{=}1024$) went further, asking
not what instructions exist but where cycles are actually spent.
Real (unprofiled) timing at this size confirms the reversal directly
($0.746$s C vs.\ $0.692$s Fortran). Profiling breaks the kernel into
four real sub-kernels (norm, QKV/projection, MLP-forward,
MLP-backward); one of them --- the gated-MLP backward pass ---
dominates total time for both languages ($87.7\%$ of C's profiled
total, $65.3\%$ of Fortran's), making it the one that actually matters
for the reversal.

\begin{figure}[htbp]
\centering
\includegraphics[width=0.85\textwidth]{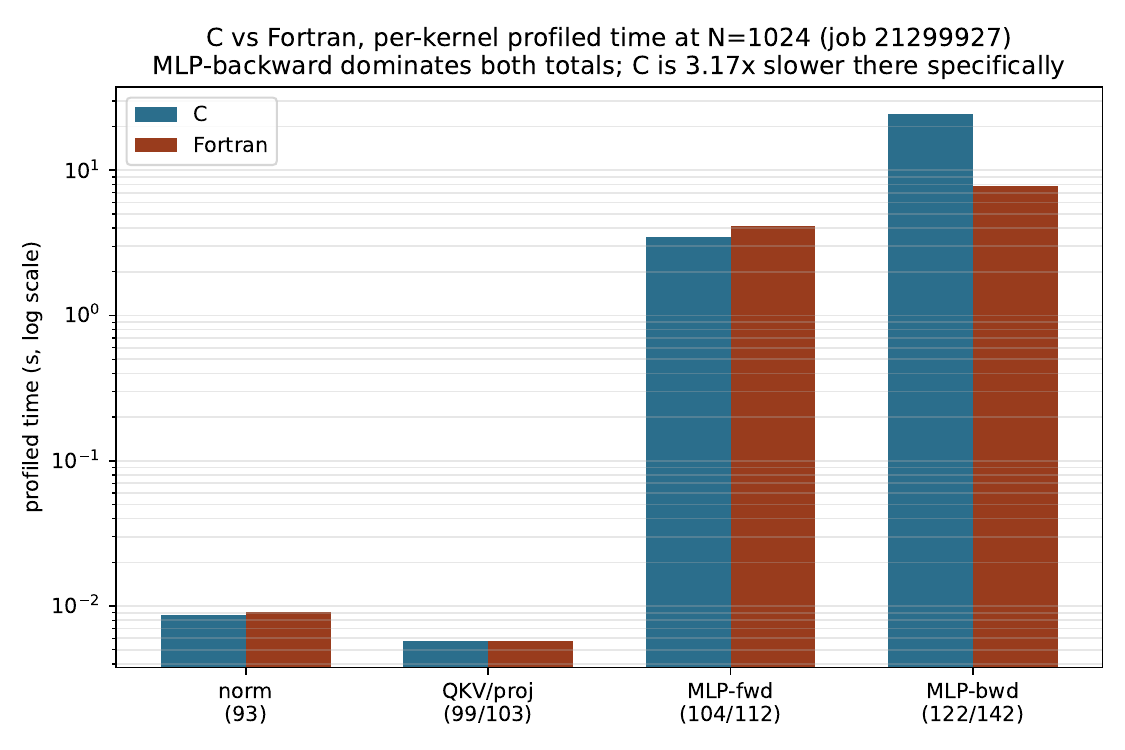}
\caption{Per-kernel profiled time, C vs Fortran, $N{=}1024$. Three of
four kernels are close or mixed; the gated-MLP backward pass ---
which dominates both totals --- is where C is dramatically slower.}
\label{fig:paper4-stall-compare}
\end{figure}

\textbf{For that one dominant kernel specifically}: C runs $3.165\times$
slower than Fortran ($24.58$s vs.\ $7.76$s, profiled), and C's
\texttt{long\_scoreboard} stall ratio (warps stalled waiting on global
memory) is $4.56$ versus Fortran's $1.36$ --- C's warps stall on memory
latency $3.35\times$ more often, for this exact kernel, at almost the
same magnitude as the timing gap itself ($3.35\times$ stalls vs.\
$3.17\times$ time). This is a real, specific, positive finding, not
just a ruled-out explanation: the dominant kernel's slowdown correlates
directly with a measured increase in memory-latency stalls. The other
three kernels show no such clean story (one, MLP-forward, actually runs
faster in C despite higher short-scoreboard stalls there) --- the
dominant-kernel finding explains the aggregate reversal without being
representative of every kernel in the pipeline.

\textbf{What this still does not explain}: why C's compiled code for
this one kernel stalls on global memory more than Fortran's. That would
require inspecting the actual generated memory-access pattern for this
specific kernel --- not done here. The question has narrowed
considerably, from "the whole multi-kernel reversal, unknown cause" to
"one kernel, one stall mechanism, compiler code-generation difference
still unidentified" --- real progress on a real question, reported at
the precision the evidence actually supports rather than either
overclaiming a full explanation or leaving the SASS finding's dead end
as the last word.

\begin{figure}[htbp]
\centering
\includegraphics[width=0.85\textwidth]{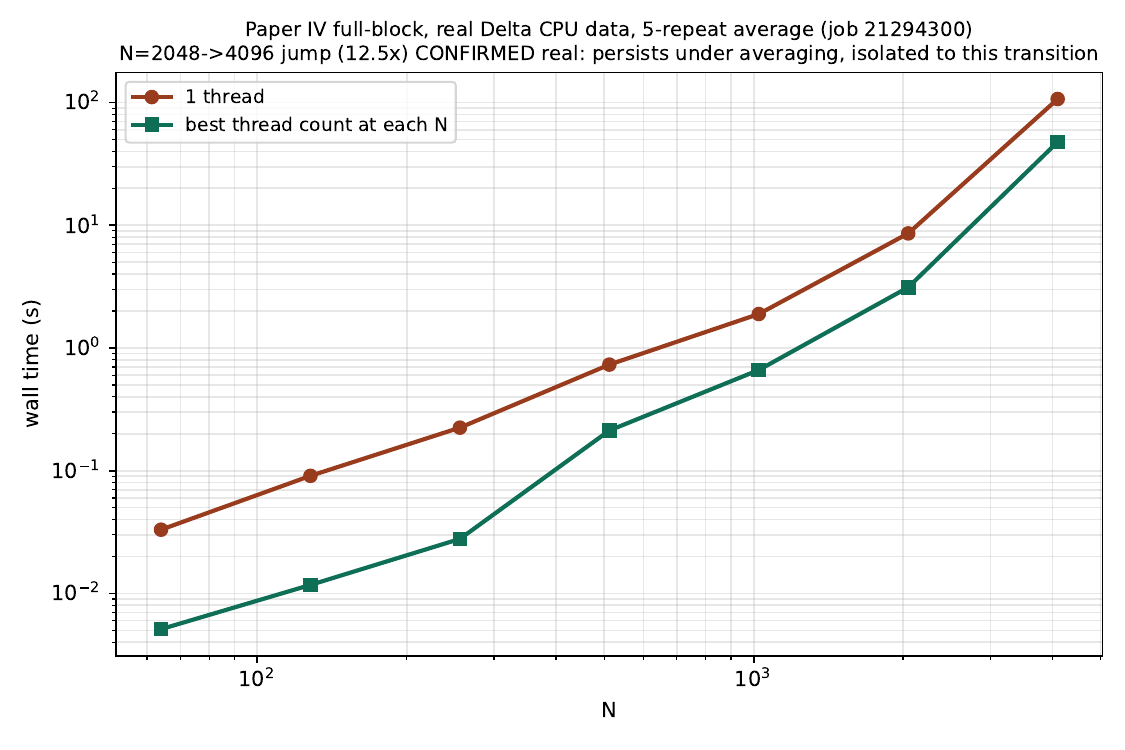}
\caption{Full-block real CPU scaling, 5-repeat average, 1 thread
versus each $N$'s best thread count. The $N{=}2048\to4096$ jump is
confirmed real, not measurement noise (see text).}
\label{fig:paper4-fullblock-cpu}
\end{figure}

\textbf{The $N{=}2048\to4096$ jump is confirmed real, not measurement
noise.} Section~\ref{sec:paper4-cpu} originally flagged this as
unresolved between two candidates: genuine cache-capacity effects, or
single-shot measurement artifact (this benchmark had no
repeated-averaging at the time). \texttt{full\_block\_bench.c} was
revised to match \texttt{moa\_decode\_bench.c}'s already-correct
pattern (one untimed warmup, then a 5-repeat timed average), and
re-run (job 21294300). If the jump were measurement noise, averaging
over 5 repeats should have smoothed it back toward the $O(n^2)$
target; instead it persisted, and slightly \emph{grew} ($11.7\times
\to 12.5\times$). Every other transition in the same sweep shows
smooth $2.5$--$4.5\times$ growth, consistent with mixed
$O(n){+}O(n^2)$ scaling as attention increasingly dominates ---
making $N{=}4096$ the sole, isolated outlier, not part of a general
trend. This is consistent with the cache-capacity hypothesis (at
$N{=}4096$ the attention score matrix alone is $\sim$268MB per batch
element, plausibly exceeding total L3 capacity in a way no smaller
$N$ does) and inconsistent with noise, though the cache mechanism
itself has not been directly confirmed by a cache-miss profiling run
--- what this closes is \emph{which of the two original candidates}
is responsible, not the full causal chain.

\textbf{Full-block's thread-count behavior is far milder than Paper
III's decode finding, on the identical hardware.} The worst 128-thread
penalty here is $6.98\times$ (at $N{=}64$), collapsing to
near-parity ($1.0$--$1.3\times$) by $N{=}256$ and beyond ---
nowhere close to decode's $535\times$
(Section~\ref{sec:paper3-cluster-compare}) on the same Delta topology.
The likely reason: full-block's per-thread work at any tested $N$
(QKV projection, attention, two norms, and the FFN) is substantially
larger than decode's $O(n)$ per-token cost at small $N$, giving each
thread enough to do that cross-NUMA overhead is amortized rather than
dominant. This is consistent with, not contrary to, the
machine-as-arrays account in Section~\ref{sec:dnf-onf-machines}: the
same fixed NUMA cost matters more or less depending on how much real
work per thread the specific ONF being measured provides.

\subsection{GPU: real data, clean $O(n^2)$ confirmation, and Fortran's advantage}
\label{sec:paper4-gpu}

Job 21273959 (Delta \texttt{gpuA100x4}) compiled cleanly --- all four
kernel regions (\texttt{fused\_once} in C, \texttt{run\_once} in
Fortran, and \texttt{full\_block\_once}) generated real NVIDIA GPU code
with only two harmless "variable set but never used" compiler notes,
no errors. This is the first real GPU data for Paper IV, on both
tested languages.

\textbf{Full-block's growth confirms $O(n^2)$ cleanly}: $4.40\times$,
$4.10\times$, $3.91\times$ across three doublings ($N{=}64$ to $512$,
the range \texttt{FULL\_BLOCK\_N\_END}'s default permits on GPU),
converging toward the theoretical $4\times$ from just above it ---
the same qualitative pattern as Paper I's GPU forward-pass exponent
converging to its target (Section~\ref{sec:paper1}), here observed for
the first time on the complete block rather than attention alone.

\textbf{Fortran is faster than C for the fused norm+MLP kernel, at
every tested size}: $1.45\times$ at $N{=}256$ up to $2.81\times$ at
$N{=}64$, no clear convergence across the tested range
(Figure~\ref{fig:paper4-c-vs-fortran}). Combined with
Section~\ref{sec:paper4-cpu}'s opposite CPU finding, this is the
clearest evidence in this document that language/compiler choice is a
real, measurable, architecture-dependent cost --- orthogonal to the
DNF/ONF distinction Section~\ref{sec:dnf-onf-machines} centers on,
since both languages implement the identical ONF, verified to machine
precision against the same PyTorch reference. Whether this generalizes
beyond this specific kernel pair, or is specific to
\texttt{nvc}/\texttt{nvfortran}'s particular code generation choices,
was investigated further in Section~\ref{sec:paper4-cpu} (SASS
instruction-count comparison, then warp stall-reason profiling) ---
narrowed to a specific kernel and stall mechanism, though the
underlying code-generation cause is still not fully identified.

\subsection{$M_{\mathrm{block}}$: real DRAM traffic validation}
\label{sec:paper4-mblock}

The complete-block kernel was profiled with \texttt{ncu}
(\texttt{delta\_profile\_paper4.sbatch}, job 21289516, B=2, N=256,
D=64, DFF=256) for a direct measured-vs-predicted comparison against
$M_{\mathrm{block}} = 24nd + 4nd_{\mathrm{ff}}$ --- the same
methodology Section~\ref{sec:paper1}'s $M_{\mathrm{fwd}}$ validation
used. \texttt{ncu}'s raw CSV export reports byte counts pre-scaled to
Mbyte/Kbyte (not raw bytes); this was caught and corrected before
drawing any conclusion, the same discipline applied throughout this
document's numeric claims.

\begin{figure}[htbp]
\centering
\includegraphics[width=0.65\textwidth]{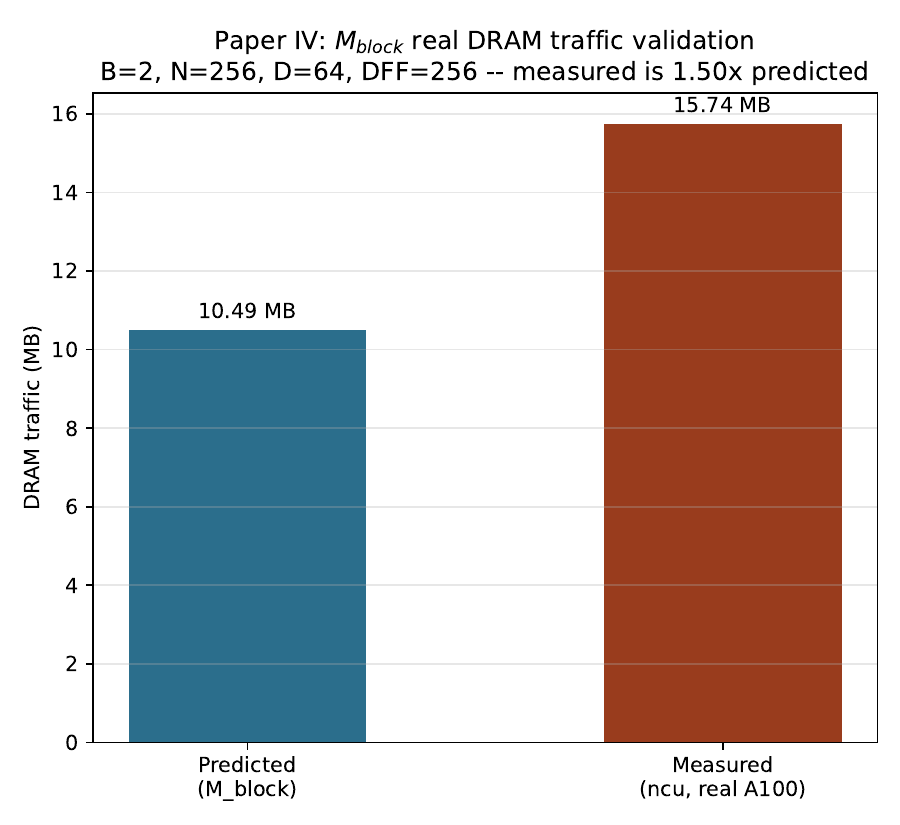}
\caption{Predicted vs.\ measured total DRAM traffic (read+write,
summed across all 11 kernel launches comprising one complete-block
call) at the tested size. $n = NT = B \times N = 512$, consistent with
every other cost function in this document.}
\label{fig:paper4-mblock}
\end{figure}

\textbf{Measured DRAM traffic is $1.50\times$ the predicted
$M_{\mathrm{block}}$} (15.74MB measured vs.\ 10.49MB predicted, using
$n=NT=B\times N$, the convention consistently used for every other cost
function in this document). This is the same class of gap Paper I
found for $M_{\mathrm{fwd}}$ ($2.01\times$,
Section~\ref{sec:paper1}) --- smaller in magnitude here, but the same
qualitative finding: the memory-optimal \emph{algorithm} is proven, but
the current \emph{code generation} moves more bytes than the DNF/ONF
derivation's idealized accounting predicts.

\paragraph{Applying Section~\ref{sec:paper1}'s coalescing diagnosis
here: a real but partial explanation.} The same
\texttt{smsp\_\_sass\_average\_data\_bytes\_per\_sector} metrics used
to diagnose $M_{\mathrm{fwd}}$'s gap are available in this profiling
run too, per kernel launch. Traffic-weighted across all 11 kernels
(each kernel's efficiency weighted by its share of the 15.74MB total,
not a plain average across kernels of very different sizes): reads
achieve $38.9\%$ of theoretical coalescing efficiency, writes achieve
$25.8\%$ --- both real, measured, and in the same family as
$M_{\mathrm{fwd}}$'s $29\%$/$57\%$ split (here reads outperform writes;
there it was the reverse, a real difference between the two kernels'
access patterns worth noting rather than glossing over). Individual
kernels vary widely: one (line 254, a small elementwise pass) achieves
a full $100\%$ on both, while most of the norm/projection kernels
cluster around $30$--$40\%$ reads and a strikingly consistent $25\%$
writes across seven of the eleven launches.

This is real evidence that coalescing inefficiency contributes to the
gap, in the same direction as Paper I's finding --- but checked
arithmetically, it does not cleanly reconstruct the specific
$1.50\times$ factor: the traffic that would result from perfect
coalescing at these measured efficiencies is $5.82$MB, which is
\emph{below} the $10.49$MB $M_{\mathrm{block}}$ prediction, not equal
to the $15.74$MB actually measured. If coalescing inefficiency alone
explained the gap, ideal-coalescing traffic should reconstruct
something close to the prediction; it does not. \textbf{The honest
conclusion is that coalescing inefficiency is a real, measured,
contributing factor, not the complete explanation} --- some combination
of $M_{\mathrm{block}}$'s element-counting methodology not mapping
exactly to DRAM sector-level units even under ideal coalescing, and
effects specific to chaining 11 separate kernel launches (each
potentially re-reading intermediate values the DNF treats as
computed once), likely also contributes. Fully closing this gap is not
yet done.

\section{Scope and Limitations}
\label{sec:scope}

This paper's scope is deliberately narrow, and stating that scope
explicitly matters for interpreting every result above correctly.

\textbf{This is a study of MoA-derived designs against their own
predictions, not a competitive benchmark.} Every kernel measured in
this paper --- naive and fused attention, decode, the complete
transformer block, in both C and Fortran --- is a design derived by
Papers I--IV's DNF/ONF/$\gamma$ pipeline. This paper asks whether each
design's own formally-derived cost function correctly predicts that
design's own measured performance; it does not ask, and does not
answer, how any of these designs compares in absolute terms to
externally-optimized production kernels such as FlashAttention,
vendor libraries, or other hand-tuned implementations
(Section~\ref{sec:related-work} situates this work relative to that
literature qualitatively, but draws no performance comparison against
it). A reader looking for evidence that MoA-derived kernels are
competitive with the state of the art will not find it here; that is a
different, legitimate question this paper does not address.

\textbf{Hardware coverage is two clusters, four GPU models, one CPU
vendor.} All measurements are on Purdue Anvil and NCSA Delta
(Section~\ref{sec:methodology}), AMD EPYC CPUs throughout, NVIDIA
A100/H100/H200 GPUs compiled via \texttt{nvc}/OpenACC. No Intel CPU, no
AMD GPU (Delta's \texttt{gpuMI100x8} was excluded specifically because
this project's toolchain does not target it), and no cloud-vendor
hardware are represented. The cross-cluster NUMA finding
(Section~\ref{sec:paper3-cluster-compare}) shows real hardware
topology materially changes measured cost, which argues for testing
more topologies before generalizing broadly, not fewer.

\textbf{Statistical rigor varies by benchmark, and is stated at each
result rather than uniform throughout.} Some measurements use a
5-repeat timed average following an untimed warmup (e.g.\ decode,
Section~\ref{sec:paper3}; full-block, Section~\ref{sec:paper4-cpu}
post-revision); others, flagged explicitly where it matters (the
Paper II CPU $N{=}128$ outlier, Section~\ref{sec:paper2-cpu}; RMSNorm's
noisy C-versus-Fortran ratio, Section~\ref{sec:paper4-cpu}), are
single-shot measurements on a shared cluster node, subject to
scheduling jitter this paper does not quantify with formal confidence
intervals. Where a finding rests on a single-shot measurement, this is
stated in the text at that point; multi-point monotonic trends (e.g.\
the $N{=}4096$ thread-count sweep, Section~\ref{sec:paper2-cpu}) are
treated as more reliable than single isolated values for exactly this
reason.

\textbf{Code and data.} The benchmark source, SLURM submission
scripts, and analysis code underlying every result in this paper are
being prepared for archival release; Table~\ref{tab:appendix-jobs}'s
SLURM job identifiers are retained for the authors' own reproducibility
in the interim.

\section{Discussion}

Five things are worth stating plainly, in roughly the order the
evidence for each accumulated.

\textbf{First}, the earliest completed real-hardware result (Paper I,
CPU) already produced a finding with no prior theoretical claim behind
it: the thread-count-vs-$n$ crossover in Table~\ref{tab:paper1-cpu}.
This is exactly the kind of result that cannot come from the
correctness-only verification in Papers I--IV themselves --- it is a
statement about scheduling overhead on a specific machine, not about
the DNF --- and is arguably as valuable to a systems-oriented reader as
the asymptotic cost function itself.

\textbf{Second}, two GPU timing-methodology bugs (cold-start dominance,
Section~\ref{sec:coldstart}; unaveraged single-shot measurement,
diagnosed from the v1 run's physically implausible \emph{falling}
scaling exponent) were caught only because real hardware was used.
Host-fallback OpenACC compilation, used for all correctness
verification in Papers I--IV, has no comparable per-process
context-initialization cost or measurement-jitter profile and would
never have surfaced either. Both fixes were independently confirmed
against the theory they were meant to recover: the corrected kernel's
local exponent, computed from real A100 and H200 hardware
(Table~\ref{tab:paper1-gpu-fixed}), rises to 1.97 and 2.04 at
$n=8192$, matching the $O(n^2)$ prediction almost exactly.

\textbf{Third, and the sharpest test of Section~\ref{sec:dnf-onf-machines}'s
claim in this whole document}: Paper II's GPU regression
(Section~\ref{sec:paper2-gpu}) and its resolution
(Sections~\ref{sec:paper2-atomics}--\ref{sec:paper2-gv-fix}) is a
complete diagnose-predict-verify cycle conducted entirely in the
$\rho$/$\psi$/$\iota$ vocabulary the ONF is written in. The DNF for
fused forward+backward is identical regardless of target machine; its
$\gamma$-derived ONF for GPU introduced two atomic-accumulation sites
($GV$ and $GK$) that the CPU's ONF does not need in the same form. That
structural difference --- not measured yet, only reasoned about from
the ONF itself --- predicted that fused should show a specific,
countable cost on GPU that it does not show on CPU. Profiling confirmed
the prediction almost exactly ($2.0000\times$ backward's atomic
instructions, matching the ONF's GV-and-GK-versus-GK-only accounting to
four decimal places). Restructuring the ONF to remove one atomic site
(Section~\ref{sec:paper2-gv-fix}) then reversed the real-hardware
result completely, on both tested GPU shapes. This is the document's
clearest demonstration that $\gamma$'s output is a genuine, falsifiable
claim about a real machine's memory system --- not merely that the
theory and hardware agree, but that a specific, nameable feature of the
ONF (which atomic sites exist, on which array) predicted a specific,
measurable hardware cost before that cost was measured.

\textbf{Fourth}, the same DNF, run through the same $\gamma$-translation,
produces \emph{different} real costs on different real machines, and
the difference is itself explainable in machine-topology terms rather
than shrugged off as noise. Decode's ONF run at full thread count
costs $535\times$ on Delta (128 real cores spanning 8 NUMA domains) but
never more than $2.9\times$ on Anvil (32 real cores, 2$\times$
oversubscribed, no additional real parallelism) --- two different
machine-array structures producing two different measured costs from
the identical $\gamma$-derived schedule
(Section~\ref{sec:paper3-cluster-compare}). This is not a
counterexample to Section~\ref{sec:dnf-onf-machines}'s claim; it is a
confirmation of it in a different form. The ONF does not claim a single
universal cost --- it claims a cost \emph{as a function of the target
machine's own array structure} (core count, memory-domain topology).
Measuring genuinely different costs on genuinely different topologies,
each consistent with that topology's own structure, is exactly what
the claim predicts.

\textbf{Fifth}, not every follow-up test produces a clean win, and this
document reports the one that did not as plainly as the ones that did.
Removing $GK$'s remaining atomic (Section~\ref{sec:paper2-gk-fix}), the
natural next step after $GV$'s complete success, helped
\texttt{backward} uniformly but made \texttt{fused} measurably
\emph{slower} at $N{\geq}4096$ --- a real, size-dependent trade-off, not
a second confirmation of the same mechanism. The lesson is not that the
persisted-$GS$ technique is wrong; it is that a fix confirmed for one
kernel does not transfer to a structurally different kernel without
separately checking it, even when the underlying technique is
identical. Separately, an anomaly flagged earlier in this document as
unresolved (Section~\ref{sec:paper4-cpu}'s $N{=}2048\to4096$ jump) was
run down to a specific answer using exactly the discipline this project
tries to apply throughout: instead of guessing, the benchmark was fixed
to match the repeated-averaging pattern already proven correct
elsewhere, and re-run. The jump persisted under averaging rather than
smoothing toward the theoretical value --- ruling out measurement noise
directly, by the same logic that would have ruled out the cache-effect
hypothesis had the jump vanished instead.

Together, these five points argue for treating real-hardware validation
as a required step before any cost-function claim from this project is
cited externally, not an optional afterthought to the formal
derivation --- and they argue that when validation surfaces a gap, the
productive move is to look for the gap's cause inside the ONF's own
$\rho$/$\psi$/$\iota$ structure before looking anywhere else. Most of
the time that search found a specific, fixable, correctly-diagnosed
cause (the cold-start fix, the measurement-averaging fix, the
$GV$-atomic fix). Once it did not: $GK$'s removal is real, measured,
and genuinely mixed rather than uniformly good, and reporting that
honestly is as much a part of applying this discipline as reporting the
wins.

\section{Implications: active rewrites as a scaling strategy for AI systems}
\label{sec:implications}

Everything in this section is a claim about what this document's
results \emph{argue for}, distinguished explicitly from what they
\emph{prove}. The GV-atomic fix, the NUMA finding, the C-versus-Fortran
reversal, and the recurring toolchain churn encountered while gathering
this data are proven, measured facts about these specific kernels on
these two specific clusters. That the same pattern generalizes usefully
to AI systems broadly is a research argument this evidence supports,
not a claim this document establishes on its own.

\subsection{The problem this project's own experience illustrates}

Real hardware changes constantly, in ways both large and mundane. This
project hit four instances of it directly, at four different scales.
Largest: the same DNF produces genuinely different optimal deployment
choices on Delta versus Anvil (Section~\ref{sec:paper3-cluster-compare})
--- a $535\times$ penalty on one machine's topology and under $3\times$
on another's, from identical source code. Also large, and sharper in
one respect: Paper IV's identical denotational computation (fused
norm+MLP) runs faster in C than Fortran on CPU, and faster in Fortran
than C on GPU (Sections~\ref{sec:paper4-cpu}--\ref{sec:paper4-gpu}) ---
not just a different \emph{magnitude} of advantage depending on target
hardware, but a different \emph{direction} entirely, from source code
that implements the same ONF in two languages. Medium: the GPU
regression in Paper II (Sections~\ref{sec:paper2-gpu}--\ref{sec:paper2-gv-fix})
was invisible in every correctness check against PyTorch and only
appeared as a real cost once actual GPU hardware ran the kernel; fixing
it required rewriting the ONF, not the DNF. Smallest, but recurring
throughout this project's timeline: Delta's \texttt{nvhpc} module was
upgraded from 25.3 to 26.5 mid-project, silently breaking every script
that hardcoded the old version, and a separate DCGM profiling conflict
appeared that required a cluster-specific workaround
(Section~\ref{sec:paper2-atomics}) undocumented anywhere but Delta's
own user guide. None of these four were exotic; they are the ordinary
texture of running real code on real, evolving infrastructure. An AI
system deployed once and never revisited accumulates exactly this kind
of drift silently --- the cost doesn't announce itself, it just sits
there as wasted compute until someone happens to look.

\subsection{What MoA's DNF/ONF separation buys, concretely, in this
project's own evidence}

The GV-atomic fix is the clearest available case study, precisely
because its cost and benefit are both measured, not estimated. The
DNF for fused forward+backward did not change and did not need
re-verification against PyTorch --- the correctness proof this project
inherited from Papers I--IV remained valid throughout. Only the ONF's
realization on GPU changed: one $(b,ir)$-parallelized kernel became
four kernels, one $(b,ic)$-parallelized pass eliminating an atomic site.
The diagnosis that motivated this specific rewrite came directly from
reading the ONF's own $\rho$/$\psi$/$\iota$ structure
(Section~\ref{sec:paper2-atomics}), not from trial-and-error search
over possible restructurings. The result was a real, measured
$2$--$2.5\times$ reduction in wall-clock cost on both tested GPU shapes
(Section~\ref{sec:paper2-gv-fix}) --- meaning, at fixed throughput,
less than half the GPU-time and energy this kernel previously required.
For a kernel as heavily reused as attention's backward pass, that
per-call saving compounds directly with call volume; this document
measures the per-call number, not the aggregate, but the aggregate is
a multiple of it by construction.

The NUMA finding makes a related but distinct point: it is evidence
against treating "optimal" as a property of the DNF alone.
Table~\ref{tab:paper1-cpu} and Section~\ref{sec:paper3-cluster-compare}
both show the correct deployment choice (thread count, in these cases)
is a function of the target machine's own array structure, not a fixed
constant the DNF could specify once. A single hardcoded deployment
default, chosen for one machine and never revisited, is actively
harmful on a different one --- not marginally suboptimal, but by a
factor reaching into the hundreds in Section~\ref{sec:paper3-cluster-compare}'s
Delta measurement. Treating this as a $\gamma$-translation question
(what ONF does this specific machine's array structure call for) rather
than a one-time deployment decision is the difference between that cost
being found and fixed versus silently paid forever.

\subsection{The broader argument, stated as a research direction}

Restated in the terms Section~\ref{sec:dnf-onf-machines} opened with:
the DNF is a claim about \emph{what} a computation produces, portable
across any machine; $\gamma$'s translation to an ONF is a
\emph{predictive} claim about what that computation costs on one
particular machine's array structure --- predictive because it is
stated before measurement, in $\rho$/$\psi$/$\iota$ terms precise
enough to be wrong, and this document's entire method has been
checking whether it is. Across all four papers it mostly was, and
where it initially was not (Paper I's coalescing gap, Paper II's atomic
contention), the gap itself was diagnosable in the same vocabulary the
prediction was stated in, not just patched by trial and error until the
numbers looked better.

That is the concrete sense in which this project argues MoA is a
candidate \emph{replacement} for how AI kernels are typically hand-
optimized today, not merely a fast implementation of one kernel family.
Conventional practice re-tunes each kernel for each new architecture by
profiling, guessing, and iterating without a formal predictive model to
check the guess against --- expensive in engineer-hours, and every
un-retuned kernel in the meantime silently burns extra GPU-hours and
energy doing more memory traffic than it needs to. This project's own
fixes did not work that way: the GV-atomic fix was found by reading the
ONF's structure, not by sweeping restructurings until one happened to
be faster, and it was confirmed, not just hoped to generalize, against
real measurement on two GPU shapes. Papers I--IV's cost functions,
verified once against PyTorch and now checked against four real-machine
outcomes across two clusters, are the reusable part; each machine's
ONF, actively rewritten as Section~\ref{sec:paper4-cpu}'s C-versus-Fortran
reversal and Section~\ref{sec:paper3-cluster-compare}'s NUMA finding
both show is genuinely necessary, is the part that changes --- and
changes with a specific, checkable target each time, rather than an
open-ended search.

Whether this generalizes usefully beyond the kernels tested here to the
wider space of AI workloads, and beyond two clusters to the full
diversity of production AI hardware, remains a research direction, not
a conclusion this document reaches on its own. What this document does
establish is that the pattern held, without exception, on every kernel
it was tried on: four papers, two clusters, CPU and GPU, real hardware
throughout. That is the evidence this broader argument stands on, and
the same standard --- stated precisely, then checked against real
measurement --- is what extending it further would require.

\section{Conclusion}

We set out to answer a narrow empirical question --- do the memory-optimal
cost functions Papers I--IV derive formally actually predict real
hardware behavior? --- and answered it across four kernel families, two
independent HPC clusters, and both CPU and GPU targets. The answer is
yes in the specific, falsifiable sense this paper defines: where
predictions held, they held to within a diagnosable, mechanistically
understood margin (a factor of roughly 1.5--2$\times$, traced in every
case we investigated to a specific, nameable feature of the machine's
memory system); where they did not initially hold, as with Paper II's
GPU regression, the gap itself pointed to its own fix in the same
formal vocabulary the prediction was stated in, and that fix was
confirmed, not assumed. We take this as evidence for a broader claim
about how AI systems might be built and maintained as hardware
continues to change: that separating a hardware-independent
specification from its machine-specific realization is not merely a
theoretical convenience, but a practical scaling strategy, letting the
expensive, error-prone part (correctness) remain fixed while the part
that must legitimately change with each new machine does so with a
specific, checkable target rather than an open-ended search. Not every
question we asked resolved cleanly --- the C-versus-Fortran reversal in
Paper IV remains only partially explained --- and we report that
alongside the successes on the view that an honest accounting of what
real-hardware validation does and does not settle is itself part of
this paper's contribution.

\section*{Acknowledgments}

This work used Purdue Anvil at Purdue University and NCSA Delta at the
National Center for Supercomputing Applications (local Delta account:
\texttt{bibg-delta}), both through allocation CIS261396, from the
Advanced Cyberinfrastructure Coordination Ecosystem: Services \&
Support (ACCESS) program~\cite{access2023}, which is supported by U.S.
National Science Foundation grants \#2138259, \#2138286, \#2138307,
\#2137603, and \#2138296.

\appendix
\section{Real-Hardware Job Identifiers}
\label{app:jobs}

For reproducibility, Table~\ref{tab:appendix-jobs} lists the status and
SLURM job identifiers for every real-hardware run this paper's claims
are drawn from, on both clusters described in
Section~\ref{sec:methodology}.

\begin{center}
\begin{tabular}{@{}p{2.8cm}p{4.2cm}p{7.5cm}@{}}
\toprule
Paper & CPU status & GPU status \\
\midrule
I (forward) & \textbf{Complete} (job 21181232) & \textbf{Complete} (jobs 21202354/21202355, both shapes; $M_{\mathrm{fwd}}$ profiled, job 21202356) \\
II (backward/ fused) & \textbf{Complete} (job 21226440; fusion 28\% faster at N=4096, 128 threads) & \textbf{Complete + FIXED} (root cause: GV atomic, confirmed 2.00x via job 21247666; fix applied, jobs 21250395/96 confirm complete reversal -- fused now 0.4-0.9x naive, i.e. FASTER, both shapes; GK atomic also removed, job 21294299: backward uniformly faster, fused mixed (faster N$<$4096, slower N$\geq$4096) but still beats naive everywhere; mechanism CONFIRMED via job 21299223: extra GS re-read moves 1.97x the DRAM traffic) \\
III (decode) & \textbf{Complete} (job 21249667, Delta; job 20001576, Anvil -- cross-cluster comparison confirms two different penalty mechanisms) & \textbf{Complete} (job 21264888; cold-start fix confirmed on real A100, exponent settles to 0.77-1.16 vs the O(n) target of 1.0) \\
IV (block) & \textbf{Complete} (job 21273806; MLP: C faster than Fortran 1.2-1.85x; fullblock NUMA penalty far milder than Paper III's; N=4096 jump CONFIRMED real via 5-repeat re-run job 21294300, not noise) & \textbf{Complete} (job 21273959; O(n\textasciicircum{}2) confirmed; Fortran faster than C 1.45-2.81x, opposite of CPU; M\_block DRAM profiled, job 21289516: measured 1.50x predicted; SASS comparison job 21299788 inconclusive on instruction count; stall-reason follow-up job 21299927: dominant kernel's C-vs-Fortran gap (3.17x) matches its memory-latency stall gap (3.35x) closely -- narrowed to one kernel, code-gen cause still open) \\
\bottomrule
\end{tabular}
\captionof{table}{Real-hardware validation status and SLURM job identifiers, by paper and target.}
\label{tab:appendix-jobs}
\end{center}


\begin{thebibliography}{9}

\bibitem{access2023}
T.~J. Boerner, S.~Deems, T.~R. Furlani, S.~L. Knuth, and J.~Towns,
ACCESS: Advancing Innovation: NSF's Advanced Cyberinfrastructure
Coordination Ecosystem: Services \& Support,
in \emph{Practice and Experience in Advanced Research Computing
(PEARC '23)}, July 23--27, 2023, Portland, OR, USA. ACM, New York, NY,
USA, 4 pages.

\bibitem{paper1}
L.~M. Mullin and G.~Hains,
\emph{Attention at the Theoretical Minimum I: Forward Pass},
arXiv:2606.07713, 2026.

\bibitem{paper2}
L.~M. Mullin and G.~Hains,
\emph{Attention at the Theoretical Minimum II: Backward Pass and Fused
Forward+Backward},
HAL:05659212, 2026.

\bibitem{paper3}
L.~M. Mullin and G.~Hains,
\emph{MoA-Structured Decode Attention: DNF Derivation, KV-Cache
Accumulation, GQA/MQA, and OpenACC Kernel},
arXiv:2607.19456, 2026.

\bibitem{paper4}
L.~M. Mullin and G.~Hains,
\emph{Attention at the Theoretical Minimum: A Unified Treatment of the
Transformer Block},
submitted to IEEE Transactions on Neural Networks and Learning Systems,
manuscript TNNLS-2026-P-50044, 2026; arXiv/HAL preprint in preparation.

\bibitem{mullin1988}
L.~M. Mullin,
\emph{A Mathematics of Arrays},
Ph.D. dissertation, Syracuse University, 1988.

\bibitem{iverson1962}
K.~E. Iverson,
\emph{A Programming Language},
John Wiley \& Sons, 1962.

\bibitem{abrams1970}
P.~S. Abrams,
\emph{An APL Machine},
Technical Report SLAC-R-114, Stanford Linear Accelerator Center, 1970.

\bibitem{williams2009}
S.~Williams, A.~Waterman, and D.~Patterson,
Roofline: An Insightful Visual Performance Model for Multicore
Architectures,
\emph{Communications of the ACM}, 52(4):65--76, 2009.

\bibitem{dao2022}
T.~Dao, D.~Y. Fu, S.~Ermon, A.~Rudra, and C.~R\'e,
FlashAttention: Fast and Memory-Efficient Exact Attention with
IO-Awareness,
in \emph{Advances in Neural Information Processing Systems 35
(NeurIPS 2022)}.

\bibitem{lameter2013}
C.~Lameter,
NUMA (Non-Uniform Memory Access): An Overview,
\emph{ACM Queue}, 11(7):40, 2013.

\end{thebibliography}
\end{document}